\documentclass[runningheads]{llncs}

\usepackage{amssymb,amsmath,amsthm,amsfonts}

\usepackage[hyphens]{url}
\usepackage{hyperref}
\usepackage[hyphenbreaks]{breakurl}
\hypersetup{
    colorlinks,
    linkcolor={red!80!black},
    citecolor={blue!80!black},
    urlcolor={blue!80!black}    
}

\usepackage{mathtools}
\usepackage{braket}
\usepackage{enumitem}
\usepackage{microtype}      
\usepackage{tikz}
\usepackage{booktabs}
\usetikzlibrary{shapes.geometric,arrows.meta,arrows,intersections,positioning,automata}
\usepackage{dsfont} 
\usepackage[lined,boxed,commentsnumbered]{algorithm2e}
\usepackage{float}
\usepackage{versions}
\usepackage{sidecap}
\usepackage[caption=false]{subfig}
\usepackage[normalem]{ulem} 
\allowdisplaybreaks

\usepackage{dsfont} 

\newcommand{\cardinality}[1]{ | #1 | }
\renewcommand{\d}[1]{\ensuremath{\operatorname{d}\!{#1}}}
\newcommand{\e}[1]{ {\mathrm{e}}^{ #1 } }
\newcommand{\im}{\mathrm{i}}

\newcommand{\expectation}[1]{ \mathbb{E} [ #1 ] }
\newcommand{\expectationWrt}[2]{ \mathbb{E}_{#2} [ #1 ] }

\newcommand{\vect}[1]{ #1 }

\newcommand{\vectOnes}[1]{ \vect{1}_{#1} }

\newcommand{\indicator}[1]{ \mathds{1} [ #1 ] }

\newcommand{\process}[2]{ \{ #1 \}_{ #2 } }

\newcommand{\pnorm}[2]{ \| #1 \|{}_{#2} }

\newcommand{\probability}[1]{ \mathbb{P} [ #1 ] }
\newcommand{\probabilityWrt}[2]{ \mathbb{P}_{#2} [ #1 ] }

\newcommand{\naturalNumbersPlus}{ \mathbb{N}_{+} }
\newcommand{\naturalNumbersZero}{ \mathbb{N}_{0} }
\newcommand{\realNumbers}{ \mathbb{R} }

\newcommand{\iterand}[2]{ #1^{[#2]} }

\newcommand{\refFigure}[1]{{\textrm{Figure~\ref{#1}}}}
\newcommand{\refTable}[1]{{\textrm{Table~\ref{#1}}}}

\newcommand{\refEquation}[1]{{\textrm{\eqref{#1}}}}

\newcommand{\refProposition}[1]{{\textrm{Proposition~\ref{#1}}}}
\newcommand{\refLemma}[1]{{\textrm{Lemma~\ref{#1}}}}

\newcommand{\refSection}[1]{{\textrm{Section~\ref{#1}}}}

\newcommand{\refAppendixSection}[1]{\S{}\textrm{\ref{#1}}}

\newcommand{\itr}[2]{ \iterand{#1}{#2} }

\def\({{\Bigl(}}
\def\){{\Bigr)}}

\newcommand{\ba}{\begin{array}}
\newcommand{\ea}{\end{array}}

\newcommand{\xdeleted}[1]{\deleted{}} 

\includeversion{abridged}
\excludeversion{unabridged}
\newcommand{\showonlyinunabridged}[1]{}
\excludeversion{abridged}
\includeversion{unabridged}
\renewcommand{\showonlyinunabridged}[1]{#1}

\usepackage[acronym]{glossaries}
\makeglossaries
\newacronym{BMC}{BMC}{Block Markov Chain}
\newacronym{LSBM}{LSBM}{Labeled Stochastic Block Model}
\newacronym{SBM}{SBM}{Stochastic Block Model}
\newacronym{SDE}{SDE}{Stochastic Differential Equation}
\newacronym{SVD}{SVD}{Singular Value Decomposition}

\def\lma#1{\textcolor{blue}{#1}}
\def\revisedPartBegin{}
\def\revisedPartEnd{}
\def\revised#1{#1}
\def\jaron#1{#1}
\def\lma#1{#1}
\begin{document}
\title{Markov chains and hitting times for error accumulation in quantum circuits}
%
%
\author{Long Ma\inst{1} \and
Jaron Sanders\inst{2}}
\authorrunning{L. Ma \& J. Sanders}
%
\institute{Delft University of Technology\\ Faculty of Electrical Engineering, Mathematics, and Computer Science\\ \email{l.ma-2@tudelft.nl}\and
Eindhoven University of Technology\\ Department of Mathematics, and Computer Science\\ \email{jaron.sanders@tue.nl}}
\maketitle 

\begin{abstract}
We study a classical model for the accumulation of errors in multi-qubit quantum computations. By modeling the error process in a quantum computation using two coupled Markov chains, we are able to capture a weak form of time-dependency between errors in the past and future. By subsequently using techniques from the field of discrete probability theory, we calculate the probability that error quantities such as the fidelity and trace distance exceed a threshold analytically. The formulae cover fairly generic error distributions, cover multi-qubit scenarios, and are applicable to e.g.\ the randomized benchmarking protocol. To combat the numerical challenge that may occur when evaluating our expressions, we additionally provide an analytical bound on the error probabilities that is of lower numerical complexity. Besides this, we study a model describing continuous errors accumulating in a single qubit. Finally, taking inspiration from the field of operations research, we illustrate how our expressions can be used to e.g.\ decide how many gates one can apply before too many errors accumulate with high probability, and how one can lower the rate of error accumulation in existing circuits through simulated annealing.
\end{abstract}

\section{Introduction}

The development of a quantum computer is expected to revolutionize computing by being able to solve hard computational problems faster than any classical computer \cite{nielsen_quantum_2011}. However, present-day state-of-the-art quantum computers are prone to errors in their calculations due to physical effects such as unwanted qubit--qubit interactions, qubit crosstalk, and state leakage \cite{preskill1998quantum}. Minor errors can be corrected, but error correction methods will still be overwhelmed once too many errors occur \cite{dgeftoler,cramer2016repeated,linke2017fault}. Quantum circuits with different numbers of qubits and circuit depths have been designed to implement algorithms more reliably \cite{fowler2004scalability}, and the susceptibility of a circuit to the accumulation of errors remains an important evaluation criterion. We therefore study now Markov chains \revised{that provide a model for} the accumulation of errors in quantum circuits. Different types of errors \cite{greenbaum2017modeling} that can occur and are included in our model are e.g.\ Pauli channels \cite{nielsen_quantum_2011}, Clifford channels \cite{magesan2013modeling,gutierrez2013approximation}, depolarizing channels \cite{nielsen_quantum_2011}, and small rotational errors \cite{bravyi2018correcting,huang2019performance}. If the random occurrence of such errors only depends on the last state of the quantum mechanical system, then the probability that error quantities such as the fidelity and trace distance accumulate beyond a threshold can be related to different hitting time distributions of two coupled Markov chains \cite{bremaud2017discrete}. These hitting time distributions are then calculated analytically using techniques from probability theory and operations research.

Error accumulation models that share similarities with the Markov chains under consideration here can primarily be found in the literature on randomized benchmarking \cite{xia2015randomized}. From the modeling point of view, the dynamical description of error accumulation that we adopt is shared in \cite{magesan2011scalable,ball2016effect,janardan2016analytical,wallman2018randomized}. These articles however do not explicitly tie the statistics of error accumulation to a hitting time analysis of a coupled Markov chain. Furthermore, while Markovianity assumptions on e.g.\ noise are common \cite{epstein2014investigating}, the explicit mention of an underlying random walk is restricted to a few papers only \cite{ball2016effect,francca2018approximate}. From the analysis point of view, research on randomized benchmarking has predominantly focused on generalizing expressions for the expected fidelity over time. For example, the expected decay rates of the fidelity are analyzed for cases of randomized benchmarking with restricted gate sets \cite{brown2018randomized}, Gaussian noise with time-correlations \cite{fong2017randomized}, gate-dependent noise \cite{wallman2018randomized}, and leakage errors \cite{wood2018quantification}; and the expected loss rate of a protocol related to randomized benchmarking is calculated in \cite{wallman2015robust,magesan2011scalable,proctor2017randomized,wallman2018randomized,carignan2018randomized}. In this article, we focus instead on the probability distributions of both the error and maximum error \revised{in the Markov chain model} -- which capture the statistics in more detail than an expectation -- for arbitrary distance measures, and in random as well as nonrandom quantum circuits. Finally, \cite{magesan2011scalable,ball2016effect,wallman2018randomized,wood2018quantification} resort to perturbation or approximate analyses (via e.g.\ Taylor expansions, and independence or decorrelation assumptions) to characterize the fidelity, whereas here we provide the exact, closed-form expressions for the distributions using the theory of Markov chains.

To be precise: this article first studies a model for discrete Markovian error accumulation in a multi-qubit quantum circuit. We suppose for simplicity that both the quantum gates and errors belong to a finite unitary group $\mathcal{G}_n \subseteq \mathcal{U}(2^{n}) $, where $ \mathcal{U}(2^{n}) $ is the unitary group for $n$ qubits. The group $\mathcal{G}_n$ can e.g.\ be the generalized Pauli group (i.e., the discrete Heisenberg--Weyl group), or the Clifford group. By modeling the quantum computation with and without errors as two coupled Markov chains living on the state space consisting of pairs of elements from these groups, we are able to capture a weak form of time-dependency within the process of error accumulation. To see this, critically note that the assumption of a Markov property does not imply that the past and the future in the quantum computation are independent given any information concerning the present \cite{bremaud2017discrete}. We must also note that while the individual elements of our two-dimensional Markov chain belong to a group, the two-dimensional Markov chain itself, here, is generally not a random walk on a group. Lastly, our Markov chain model works for an arbitrary number of qubits. These model features are all relevant to the topic of error modeling in quantum computing, and since the Markov property is satisfied in randomized benchmarking, the model has immediate application. \jaron{The method is generic in the sense that any measure of distance between two pure quantum states may be used to quantify the error, and that it allows for a wide range of error distributions. The method can handle nonuniform, gate-, and time-dependent errors. Concretely, for arbitrary measure of distance and a wide range of error distributions, we will calculate (i) the expected error at time $t$, (ii) the probability that an error is larger than a threshold $\delta$ at time $t$, and (iii) the probability that the error has \emph{ever} been larger than a threshold $\delta$ before time $t$, and we do so both for random and nonrandom circuits.}

In addition to studying a model for discrete Markovian error accumulation in quantum circuits, we also briefly study a random walk model on the three-dimensional sphere \cite{roberts1960random}. This model is commonly used to describe the average dephasing of a single qubit (or spin) \cite{gutmann2005description}. Using this model, we characterize the distribution and expectation of the trace distance measuring the error that is accumulated over time. These derivations are, essentially, refinements that provide information about the higher-order statistics of the error accumulation in a single qubit.

The approach taken in this article is a hybrid between classical probability theory and quantum information theory. This hybridization allows us to do quite detailed calculations, but not every quantum channel will satisfy the necessary assumptions such as Markovianity of the error distribution. On the other hand, in cases where one introduces their own source of randomness (such as in randomized benchmarking), the assumptions are met naturally. 
It should furthermore be noted that the numerical complexity of the exact expressions \revised{we provide} is high for large quantum circuits. The precise difficulty of evaluating our expressions depends on the particulars of the quantum circuit one looks at. For practical purposes, we therefore also provide an explicit bound on the maximum error probability that is of lower numerical complexity. Furthermore, we also discuss a reduction in complexity that occurs when starting a quantum computation from a stabilizer state: the coupled Markov chain's state space then reduces in size. Reference \cite{janardan2016analytical} is relevant to mention here, because similar to our observations, these authors also note the generally high computational complexity of error analysis in quantum circuits. The issue is approached in \cite{janardan2016analytical} differently and in fact combinatorially by converting circuits into directed graphs, tracing so-called fault-paths through these graphs, and therewith estimating the success rates of circuits. 

To illustrate and substantiate our theoretical results, we provide detailed discussions of further numerical experiments that we ran with a quantum simulator purpose-built for this research. Experiments include: the application of our formulae to randomized benchmarking; a comparison between simulated results on the accumulation of errors on a single qubit and our explicit formulae, \jaron{as well as to a traditional method that calculates the evolution of the trace distance when repeating a depolarization channel}; the effect of gate-dependent error distributions on the accumulation of errors in a one- and two-dimensional quantum circuit; \jaron{and lowering the misclassification probability in a circuit that implements the Deutsch--Jozsa Algorithm for one classical bit}. 

Finally, we use the expressions that describe how likely it is that errors accumulate to answer two operational questions that will help advance the domain of practical quantum computing \cite{knill2005quantum}. First, we calculate and bound analytically how many quantum gates $t^\star_{\delta,\gamma}$ one can apply before an error measure of your choice exceeds a threshold $\delta$ with a probability above $\gamma$. This information is useful for e.g.\ deciding how often a quantum computer should perform repairs on qubits, and is particularly opportune at this moment since quantum gates fail $O(0.1$--$1\%)$ of the time \cite{knill2005quantum}. Related but different ideas can be found in e.g.\ \cite[\S{2.3}]{greenbaum2017modeling}, where the accumulation of bit-flips and rotations on a repetition code is studied and a time to failure is derived, and in \cite[\S{V}]{harper2019statistical}, where an upper bound on the number of necessary measurements for a randomized benchmarking protocol is derived. Second, using techniques from optimization, we design a simulated annealing method that improves existing circuits by swapping out gate pairs to achieve lower rates of error accumulation. There is related literature where the aim is to reduce the circuit depth \cite{maslov2008quantum,kliuchnikov2013optimization,amy2019formal}, but an explicit expression for error accumulation has not yet been leveraged in the same way. Moreover, we also discuss conditions under which this tailor-made method is guaranteed to find the best possible circuit. Both of these excursions illustrate how the availability of an analytical expression for the accumulation of errors allows us to proceed with second-tier optimization methods to facilitate quantum computers in the long-term. We further offer an additional proof-of-concept that simulated annealing algorithms can reduce error accumulation rates in existing quantum circuits when taking error distributions into account: \jaron{we illustrate that the misclassification probability in a circuit that implements the Deutsch--Jozsa Algorithm for one classical bit \cite{deutsch1992rapid,cleve1998quantum} can be lowered by over $40\%$. In this proof of concept we have chosen an example error distribution that is gate-dependent and moreover one that is such that \emph{not} applying a gate gives the lowest error rate in this model; applying a single-qubit gate results in a medium error rate; and applying a two-qubit gate gives the largest probability that an error may occur.}

This paper is structured as follows. In \refSection{sec:Model}, we give the model aspects pertaining to the quantum computation (gates, error dynamics, and error measures) and we introduce the coupled Markov chain \revised{to describe} error accumulation. In \refSection{sec:Error_accumulation}, we provide the relation between the probability of error and the hitting time distributions, and we derive the error distributions as well as its bound. We also calculate the higher-order statistics of an error accumulation model for a single qubit that undergoes (continuous) random phase kicks and depolarization. In \refSection{sec:Numerical_results}, we illustrate our theoretical results by comparing to numerical results of a quantum simulator we wrote for this article. In \refSection{sec:Circuit_optimization}, we discuss the simulated annealing scheme. Finally, in \refSection{sec:Conclusion}, we conclude with ideas for future research.
\section{Model and coupled Markov chain}
\label{sec:Model}
\subsection{Gates and errors in quantum computing}

It is generally difficult to describe large quantum systems on a classical computer for the reason that the state space required increases exponentially in size with the number of qubits \cite{moll2018quantum}. However, the stabilizer formalism is an efficient tool to analyze such complex systems \cite{fujii2015stabilizer}. Moreover, the stabilizer formalism covers many paradoxes in quantum mechanics \cite{aaronson2004improved}, including the Greenberger--Horne--Zeilinger (GHZ) experiment \cite{greenberger1989going}, dense quantum coding \cite{bennett1992communication}, and quantum teleportation \cite{bennett1993teleporting}. Specifically, the stabilizer circuits are the smallest class of quantum circuits that consist of the following four gates:
$
\omega = \e{\im \pi /4 }
$, 
$
H
= ( {1} / {\sqrt{2}} )
\bigl(
   ( 1, 1 ); \allowbreak
   ( 1,-1 )
\bigr)
$,  
$
S
= 
\bigl( 
   ( 1, 0  ); \allowbreak
   ( 0, \im )
\bigr)
$,
and
$
   Z_{c}
   =
   \bigl(
      (1, 0, 0, 0); \allowbreak
      (0, 1, 0, 0); \allowbreak
      (0, 0, 1, 0); \allowbreak
      (0, 0, 0,-1)
   \bigr).   
$
These four gates are closed under the operations of tensor product and composition \cite{Selinger2013GeneratorsAR}. As a consequence of the Gottesman--Knill theorem, stabilizer circuits can be efficiently simulated on a classical computer \cite{gottesman1998heisenberg}.

Unitary stabilizer circuits are also known as the Clifford circuits; the Clifford group $\mathcal{C}_n$ can be defined as follows. First: let $P \triangleq \{ I, X, Y, Z \}$ denote the Pauli matrices, so 
$
I
=
\left( 
( 1, 0 );
( 0, 1 )
\right)
$, 
$
X
=
\left( 
( 0, 1 );
( 1, 0 ) 
\right)
$, 
$
Y
=
\left( 
( 0,-i );
( i, 0 )
\right)
$, 
and 
$
Z
=
\left( 
( 1, 0 );
( 0,-1 )
\right)
$,
and let 
$
P_n
\triangleq \bigl\{ \sigma_1 \otimes \cdots \otimes \sigma_n \mid \sigma_i \in P \bigr\}
$
denote the Pauli matrices on $n$ qubits. The Pauli matrices are commonly used to model errors that can occur due to the interactions of the qubit with its environment \cite{ruskai2000pauli}. In the case of a single qubit, the matrix $ I $ represents that there is no error, the matrix $ X $ that there is a bit-flip error, the matrix $ Z $ that there is a phase-flip error, and the matrix $ Y $ that there are both a bit-flip and a phase-flip error. The multi-qubit case interpretations follow analogously. Second: let $P_n^* = P_n / I^{ \otimes n }$. We now define the Clifford group on $n$ qubits by
$
\mathcal{C}_n 
\triangleq \bigl\{ U \in  \mathcal{U}(2^n) \mid\sigma \in \pm P_n^* \Rightarrow U \sigma U^\dagger \in \pm P_n^* \bigr\} / \mathcal{U}(1)
$.

The fact that $\mathcal{C}_n$ is a group can be verified by checking the two necessary properties (see \refAppendixSection{appendix:group}). The Clifford group on $ n $ qubits is finite \cite{koenig2014efficiently}, \jaron{and we will ignore the global phase throughout this paper for convenience;} its size is then
$
\left|\mathcal{C}_n\right| = 2^{n^{2}+2n}\mathop \prod_{i = 1}^{n} \left( 4^{i}-1\right).
$
Moreover, for a single qubit, a representation for the Clifford group $ {\mathcal{C}}_{1}= \lbrace  C_{1},C_{2}, \cdots,C_{24} \rbrace $ can then be enumerated and its elements are for example shown in \cite{xia2015randomized} and \cite{ball2016effect}.

\subsection{Dynamics of error accumulation}

Suppose that we had a faultless, perfect quantum computer. Then a faultless quantum mechanical state $ \rho_{t} $ at time $t$ could be calculated under a gate sequence $ \mathcal{U}_\tau= \lbrace U_{1},\ldots,U_{\tau} \rbrace $ from the initial state $ \rho_{0}\triangleq \ket{\psi_{0}}\bra{\psi_{0}} $. Here $ \tau<\infty $ denotes the sequence length, and $ t\in\{0, 1, \cdots, \tau\} $ enumerates the intermediate steps. On the other hand, with an imperfect quantum computer, a possibly faulty quantum mechanical state $ \sigma_{t} $ at time $t$ would be calculated under both $ \mathcal{U}_t $ and some (unknown) noise sequence $ \mathcal{E}_t= \lbrace \Lambda_{1}, \ldots, \Lambda_{t} \rbrace $ starting from an initial state $\sigma_{0} \triangleq \ket{\Psi_{0}} \bra{\Psi_{0}}$ possibly different from $\rho_0$. We define the set of all pure states for $n$ qubits as $ \mathcal{S}^n $ and consider the situation that $ \ket{\psi_0}, \ket{\Psi_0} \in \mathcal{S}^n$.

To be precise, define for the faultless quantum computation \begin{equation}
\rho_{t}
\triangleq \ket{ \psi_{t} } \bra{ \psi_{t} }=U_{t}\ket{ \psi_{t-1} } \bra{ \psi_{t-1} }{U^{\dagger}_{t}}
\end{equation} 
for times $ t=1,2, \ldots ,\tau $. Let $ X_t\triangleq U_t U_{t-1}\cdots U_1 $ be shorthand notation such that $ \rho_{t}=X_t \rho_{0} {X^{\dagger}_t}$. For the possibly faulty quantum computation, define
\begin{equation*}
\sigma_{t}
\triangleq \ket{ \Psi_{t} } \bra{ \Psi_{t} }=\Lambda_{t}U_{t}\ket{ \Psi_{t-1} } \bra{ \Psi_{t-1} }{U^{\dagger}_{t}}{\Lambda^{\dagger}_{t}}
\end{equation*}
for times $ t=1, 2, \ldots ,\tau $, respectively. Introduce also the shorthand notation $ Y_t\triangleq \Lambda_t U_t \Lambda_{t-1} U_{t-1}\cdots \Lambda_1 U_1 $ such that $ \sigma_{t}=Y_t \sigma_{0} {Y^{\dagger}_t}$. The analysis in this paper can immediately be extended to the case where errors (also) precede the gate. The error accumulation process is also illustrated in \refFigure{fig:Quantum_mechanical_state_driving}.

\begin{figure}[!hbtp]
\vspace{-0.25cm}
\centering
\begin{tikzpicture}[
every node/.style={node distance=2.5cm},
gate/.style={rectangle, draw, minimum height=0.5cm, minimum width=0.5cm},
state/.style={rectangle, minimum height=0.85cm, minimum width=0.5cm}
]
\node [state] (C) {a)};
\node [state, right=0cm of C] (G) {Faultless computation:};
;
\node [state, below=0cm of C] (A) {$\rho_0$};
\node [state, right=1.2cm of A] (B) {$\rho_1$};
\node [state, draw=none, right=1.2cm of B] (D) {$\ldots$};
\node [state,  right=1.4cm of D] (E) {$\rho_{\tau-1}$};
\node [state, right=1.2cm of E] (F) {$\rho_\tau$};
\path[->,thick, every node/.style={anchor=south}]
(A) edge node {$U_1$} (B)
(B) edge node {$U_2$} (D)
(D) edge node {$U_{\tau-1}$} (E)
(E) edge node {$U_{\tau}$} (F)
;
\node [state,below=1cm of C] (C2) {b)};
\node [state, right=0cm of C2] (G2){Potentially faulty computation:};
;
\node [state, below=0cm of C2] (A2) {$\sigma_0$};
\node [state, right=1.2cm of A2] (B2) {$\sigma_1$};
\node [state, draw=none, right=1.2cm of B2] (D2) {$\ldots$};
\node [state,  right=1.4cm of D2] (E2) {$\sigma_{\tau-1}$};
\node [state, right=1.2cm of E2] (F2) {$\sigma_\tau $};

\path[->,thick, every node/.style={anchor=south}]
(A2) edge node {$ \Lambda_1U_1$} (B2)
(B2) edge node {$\Lambda_2U_2$} (D2)
(D2) edge node {$\Lambda_{\tau-1}U_{\tau-1}$} (E2)
(E2) edge node {$\Lambda_{\tau}U_\tau$} (F2)
;
\end{tikzpicture} 
\caption{Schematic depiction of the coupled quantum mechanical states $ \rho_{t} $ and $ \sigma_{t} $ for times $ t=0, 1, \cdots, \tau $. a) Faultless computation. The state $\rho_{t}$ is calculated based on a gate sequence $ \mathcal{U}_t= \lbrace U_{1},\ldots,U_{t} \rbrace $ from the initial state $\rho_{0}$. b) Potentially faulty computation. The state $\sigma_{t}$ is calculated using \emph{the same} gate sequence $ \mathcal{U}_t= \lbrace U_{1},\ldots,U_{t} \rbrace $ and an additional error sequence $ \mathcal{E}_t= \lbrace \Lambda_{1},\ldots,\Lambda_{t} \rbrace $. The final state $\sigma_{\tau}$ can depart from the faultless state $\rho_{\tau}$ because of errors.}
\label{fig:Quantum_mechanical_state_driving}
\vspace{-0.25cm}
\end{figure}

\subsection{Distance measures for quantum errors}

The error can be quantified by any measure of distance between the faultless quantum-mechanical state $ \rho_t $ and the possibly faulty quantum-mechanical state $ \sigma_t $ for steps $ t=0, 1, \ldots, \tau $. For example, we can use the fidelity  
$
F_t\triangleq \text{Tr}\sqrt{{{\rho}_{t}}^{1/2}\sigma_{t}{{\rho}_{t}}^{1/2}}
$ \cite{nielsen_quantum_2011},
or the Schatten $d$-norm \cite{bhatia2013matrix} defined by
\begin{equation}
D_t
\triangleq \pnorm{ \sigma_t - \rho_t }{d} 
= \tfrac{1}{2} \text{Tr}{\left[  \left\lbrace  ( \sigma_t - \rho_t )^\dagger ( \sigma_t - \rho_t ) \right\rbrace ^{ \frac{d}{2} } \right] }^{ \frac{1}{d} }
\end{equation}
for any $d \in [1,\infty)$. The Schatten $ d $--norm reduces to the trace distance for $d = 1$, the Frobenius norm for $d = 2$, and the spectral norm for $d = \infty$. In the case of one qubit, the trace distance between quantum-mechanical states $\rho_t$ and $\sigma_t$ equals half of the Euclidean distance between $\rho_{t}$ and $\sigma_{t}$ when representing them on the Bloch sphere \cite{nielsen_quantum_2011}. It is well known that the trace distance is invariant under unitary transformations \cite{nielsen_quantum_2011}; a fact that we leverage in \refSection{sec:Error_accumulation}.

\begin{figure}[!hbtp]
\centering
\begin{tikzpicture}[->,>=stealth',shorten >=1pt,auto,node distance=2.7cm,semithick]
  \tikzstyle{every state}=[fill=none,draw=black,text=black, minimum size=1.1cm]

  \node[state] (A)                    {$\rho_0= \sigma_0$};
  \node[state]         (B) [above right of=A] {$\rho_1= \sigma_1$};
  \node[state]         (C) [right of=B] {$\rho_2= \sigma_2$};
  \node[state]         (D) [below right of=C] {$\rho_3$};
  \node[state]         (E) [below left of=D, draw=none]       {$\cdots$};
  \node[state]         (F) [left of=E] {$\sigma_3$};
  
\path 	(A) edge [bend left=10] node {$\Lambda_1 = I^{ \otimes n }$, $\Lambda_1U_1 $} (B)
		(B)	edge [bend left=30] node {$\Lambda_2 = I^{ \otimes n }$, $\Lambda_2U_2 $} (C)
		(C)	edge [bend left=10] node {$U_3$} (D)
		;
\path 	(A) edge [bend right=10] node [below right] {$U_1$} (B)
		(B)	edge [bend right=10] node [below] {$U_2$} (C)
		(C)	edge [bend right=10] node {$\Lambda_3 \neq I^{ \otimes n }$, $\Lambda_3U_3 $} (F)		
		;
\end{tikzpicture}
\caption{Coupled chain describing the quantum circuit with errors. In this depiction, we start from \emph{the same} initial state for simplicity. Here an error $\Lambda_3\neq I^{ \otimes n }$ occurs as the third gate is applied. Note that the coupled chain $\rho_t$, $\sigma_t$ separates.}
\label{fig:coupled_random_walk}
\end{figure}
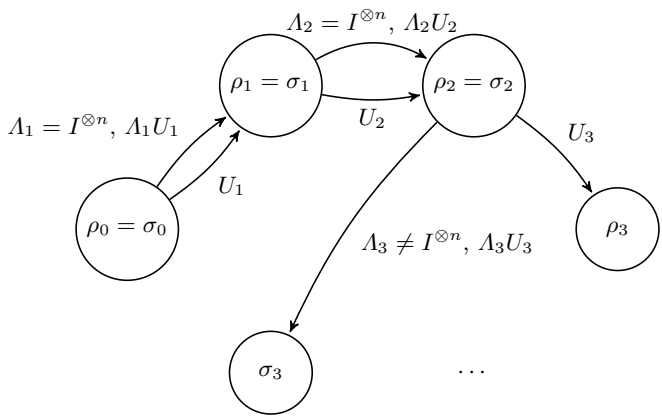

In this paper, we are going to analyze the statistical properties of some arbitrary distance measure (one may choose) between the quantum mechanical states $ \rho_{t} $ and $ \sigma_{t} $ for times $ t=0, 1, \ldots, \tau $. For illustration, we will state the results in terms of the Schatten \lma{$ d $}--norm, and so are after its expectation $\expectation{D_t}$, as well as its distributions $\probability{ D_t \leq \delta }$, $\probability{ \max_{0 \leq s \leq t} D_s \leq \delta }$. Throughout this paper, the operator $\mathbb{P}$ and thus also $\mathbb{E}$ are with respect to a sufficiently rich probability space $(\Omega, \mathbb{P}, \mathcal{F})$ that each time can describe the Markov chain being considered. 
As we show in \refAppendixSection{appendix:Relation_between_the_error_probabilities_when_using_the_trace_distance_and_fidelity}, in case of the trace distance (\lma{$d = 1$}), these probabilities can then be related to the corresponding probabilities for the fidelity:
\begin{lemma}
\label{lem:Relation_between_PFt_and_PDt}
With \lma{$ d=1 $}, it holds that $\probability{ F_t \geq 1 - \varepsilon } \geq \probability{ D_t \leq \varepsilon }$ for all $t \geq 0$. Furthermore,
\begin{equation}
\probability{ \min_{0 \leq s \leq t} F_s \geq 1 - \varepsilon } \geq \probability{ \max_{0 \leq s \leq t} D_s \leq \varepsilon }.
\end{equation}
\end{lemma}
\section{Error accumulation}
\label{sec:Error_accumulation}

\subsection{Discrete, random error accumulation (multi-qubit case)}
\label{sec:Discrete_random_error_accumulation}

Following the model described in \refSection{sec:Model} and illustrated in Figure~\ref{fig:Quantum_mechanical_state_driving} and Figure~\ref{fig:coupled_random_walk}, we define the gate pairs $ Z_t\triangleq(X_t, Y_t) $ for $t= 0, 1, 2, \ldots, \tau $, and suppose that $Z_0 = z_0$ with probability one where $z_0 = (x_0, y_0)$ is deterministic and given a priori. Note in particular that if the initial state is prepared without error, then $ \rho_0=\sigma_0 $ and consequently $z_0=(I^{ \otimes n }, I^{ \otimes n }) $. If on the other hand the initial state is e.g.\ prepared incorrectly as $y_0 \ket{ \psi_0 }$ instead of $\ket{\psi_0}$, then $z_0 = ( I^{ \otimes n }, y_0 )$.

\subsubsection{The case of random circuits}
\label{sec:The_case_of_random_circuits}

We consider first the scenario that each next gate is selected randomly and independently from everything but the last system state. This assumption is satisfied in e.g.\ the randomized benchmarking protocol \cite{xia2015randomized,magesan2011scalable,ball2016effect,janardan2016analytical,wallman2018randomized,epstein2014investigating,francca2018approximate,brown2018randomized,fong2017randomized,wood2018quantification,wallman2015robust}. The probabilities $ \probabilityWrt{ D_t > \delta }{z_0} $ and $ \probabilityWrt{\max_{0 \leq s \leq t} D_s \leq \delta}{z_0} $ can then be calculated once the initial states $\ket{\psi_0}$, $\ket{\Psi_0}$ and the \emph{transition matrix} are known. Here, the subscript $z_0$ reminds us of the initial state the Markov chain is started from.

Let the transition matrix of the Markov chain $\process{ Z_t }{t \geq 0}$ be denoted element-wise by
$
P_{z,w} 
\triangleq \probability{ Z_{t+1} = w | Z_t = z }
$
for $z = (x,y), w = (u,v) \in \mathcal{G}_n^2$. The transition matrix satisfies $ P\in {[0, 1]}^{{|\mathcal{G}_n |}^2\times {|\mathcal{G}_n |}^2} $ and the elements of each of its rows sum to one. Let
\begin{equation}
    P_{z_0,w}^{(t)} 
    \triangleq 
    \probability{ Z_t = w | Z_0 = z_0 } 
    = ( P^t )_{z_0,w}
     \label{eqn:Pzwt_for_homogeneous_Markov_chain}
\end{equation}
stand in for the probability that the process is at state $w$ at time $t$ starting from $Z_0 = z_0$. Note that the second equality follows from the Markov property \cite{bremaud2017discrete}. 

\emph{Example 1:} Consider the situation that the error depends on the last gate. The transition probability $ P_{z,w} $ for $z = (x,y), w = (u,v) \in \mathcal{G}_n^2$ can then be calculated as follows. For the faultless computation, a gate $ U=ux^{-1} $ that transfers the density matrix $ x\rho_0 x^\dagger $ to $ u\rho_0 u^\dagger$ is randomly chosen according to a gate probability vector $\kappa$. For the possibly faulty computation, an error that transfers the density matrix $ y\sigma_0 y^\dagger $ to $ v\sigma_0 v^\dagger  $, after the gate $ U=ux^{-1} $, is $ \Lambda=vy^{-1}xu^{-1} $. Let $\zeta(\Lambda=vy^{-1}xu^{-1} |ux^{-1})$ denote the probability that the error $ \Lambda=vy^{-1}xu^{-1}$ occurs given that the gate $ U=ux^{-1} $ just occurred. The transition matrix then satisfies $ \probability{ Z_{t+1} = w | Z_t = z }= \kappa(U=ux^{-1})\zeta(\Lambda=vy^{-1}xu^{-1} |ux^{-1})$ component-wise.

\emph{Example 2:} If we assume that errors and gates are independently generated, then the transition matrix satisfies $ \probability{ Z_{t+1} = w | Z_t = z }=\kappa(U=ux^{-1})\zeta(\Lambda=vy^{-1}xu^{-1}) $ component-wise.

\begin{figure}[!hbtp]
\centering
\includegraphics[
width=0.5\linewidth
]%
{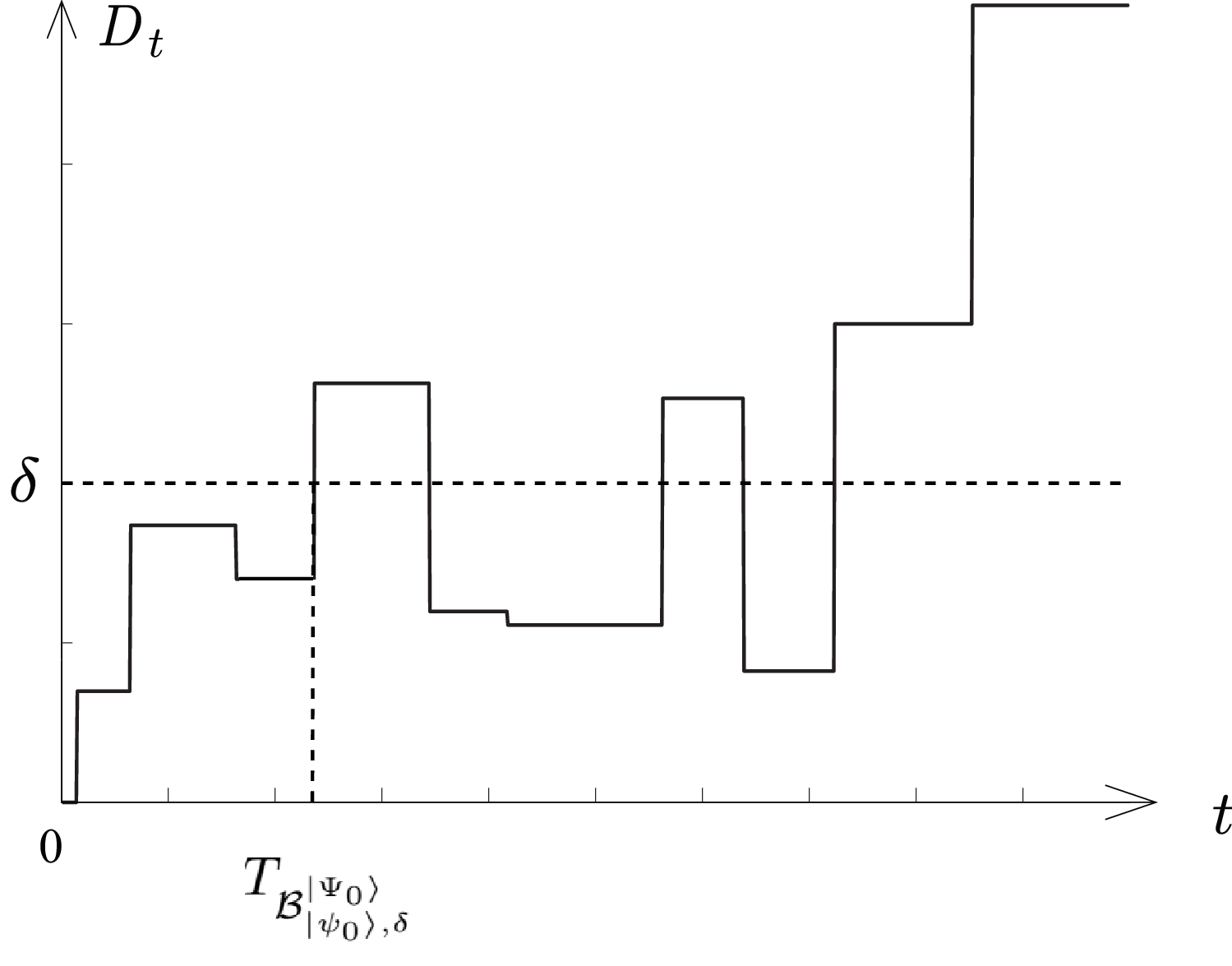}%
\caption{Schematic diagram of the hitting time $ T_{\mathcal{B}_{\ket{\psi_0},\delta}^{\ket{\Psi_0}}}  $.}%
\label{fig:Schematic_depiction_of_the_hitting_time}%
\end{figure}

We are now after the probability that the distance $ D_t $ is larger than a threshold $ \delta $. We define thereto the set of \emph{$\delta$-bad gate pairs} by
\begin{equation}
\mathcal{B}_{\ket{\psi_0},\delta}^{\ket{\Psi_0}}\triangleq \bigl\{ (x,y) \in \mathcal{G}_{n}^2 \big| \pnorm{ x \rho_0 x^\dagger - y \sigma_0 y^\dagger }{d} > \delta \bigr\}
\label{eqn:Definition_of_the_set_of_delta_bad_gate_pairs}
\end{equation}
for $ \ket{\psi_0}, \ket{\Psi_0} \in \mathcal{S}^n, \delta \geq 0 $, as well as the \emph{hitting time} of any set $\mathcal{A} \subseteq \mathcal{G}_{n}^2$ by 
\begin{equation}
T_{\mathcal{A}}
\triangleq \inf \{ t \geq 0 | Z_t \in \mathcal{A} \}
\label{eqn:Definition_of_a_hitting_time}
\end{equation}
with the convention that $\inf \phi = \infty$. Note that $T_{\mathcal{A}} \in \naturalNumbersZero \cup \{ \infty \}$ and that it is random. With definitions \eqref{eqn:Definition_of_the_set_of_delta_bad_gate_pairs}, \eqref{eqn:Definition_of_a_hitting_time}, we have the convenient representation
\begin{equation}
\probabilityWrt{ \max_{0 \leq s \leq t} D_s \leq \delta }{z_0}
= 1 - \probabilityWrt{ \max_{0 \leq s \leq t} D_s > \delta }{z_0}
= 1 - \probabilityWrt{ T_{\mathcal{B}_{\ket{\psi_0},\delta}^{\ket{\Psi_0}}} \leq t }{z_0}
\label{eqn:Probability_of_maximum_error_written_as_a_hitting_time}
\end{equation}
for this homogeneous Markov chain. As a consequence of \eqref{eqn:Probability_of_maximum_error_written_as_a_hitting_time}, the analysis comes down to an analysis of the hitting time distribution for this coupled Markov chain (\refFigure{fig:Schematic_depiction_of_the_hitting_time}).

\paragraph{Results.} 

Define the matrix 
$
B_{\ket{\psi_0},\delta}^{\ket{\Psi_0}} \in [0,1]^{\cardinality{\mathcal{G}_n }^2\times\cardinality{\mathcal{G}_n }^2}
$ 
element-wise by
\begin{equation}
\bigl( B_{\ket{\psi_0},\delta}^{\ket{\Psi_0}} \bigr)_{z,w}
\triangleq
\begin{cases}
P_{z,w} & \textrm{if} \quad w \not\in \mathcal{B}_{\ket{\psi_0},\delta}^{\ket{\Psi_0}}, \\
0 & \textrm{otherwise}.
\end{cases}
\label{eqn:Definition_of_the_B_matrix_for_random_circuits}
\end{equation}
Let the initial state vector be denoted by $ e_{z_0} $, a $\cardinality{\mathcal{G}_n }^2\times 1$ vector with just the $z_0$-th element $ 1 $ and the others $0$. Also let $\vectOnes{ \mathcal{A} }$ denote the $ \cardinality{\mathcal{G}_n }^2\times 1$ vector with ones in every coordinate corresponding to an element in the set $\mathcal{A}$. Let the transpose of an arbitrary matrix $A$ be denoted by $A^{\mathrm{T}}$ and defined element-wise $( A^{\mathrm{T}} )_{i,j} = A_{j,i}$. Finally, we define a $ \cardinality{\mathcal{G}_n }^2\times 1$ vector
$
d_{\ket{\psi_0}}^{\ket{\Psi_0}}
= \bigl( \pnorm{ x \rho_0 x^\dagger - y \sigma_0 y^\dagger }{d} \bigr)_{ (x,y) \in \mathcal{G}^2_n }
$
enumerating all possible Schatten $d$-norm distances. We now state our first result:

\begin{proposition}[Error accumulation in random circuits]
\label{prop:Error_accumulation_distributions_for_random_circuits}
For any $ z_0\in \mathcal{G}_n^2 $, $\delta \geq 0$, $t=0, 1, \ldots, \tau<\infty$: the expected error is given by
\begin{equation}
\expectationWrt{D_t} {z_0}
=e_{z_0}^\mathrm{T} P^t{d_{\ket{\psi_0}}^{\ket{\Psi_0}}}.
\label{eqn:Random_circuits__expectation}
\end{equation}
Similarly,
the distribution of error is given by
\begin{equation}
\probabilityWrt{ D_t > \delta }{z_0}
= e_{z_0}^\mathrm{T} P^t\vectOnes{ \mathcal{B}_{\ket{\psi_0},\delta}^{\ket{\Psi_0}} },
\label{eqn:Random_circuits__Distribution}
\end{equation}
and is nonincreasing in $\delta$. 
Furthermore; if $z_0 \not\in \mathcal{B}_{\ket{\psi_0},\delta}^{\ket{\Psi_0}}$, the distribution of maximum error is given by
\begin{equation}
\probabilityWrt{ \max_{0 \leq s \leq t} D_s > \delta }{z_0} 
= \sum\limits_{s=1}^{t} e_{z_0}^\mathrm{T} \bigl( B_{\ket{\psi_0},\delta}^{\ket{\Psi_0}} \bigr)^{s-1} \bigl( P - B_{\ket{\psi_0},\delta}^{\ket{\Psi_0}} \bigr) \vectOnes{ \mathcal{B}_{\ket{\psi_0},\delta}^{\ket{\Psi_0}} }, 
\label{eqn:Random_circuits__Maximum}
\end{equation}
and otherwise it equals one. Lastly, \eqref{eqn:Random_circuits__Maximum} is  nonincreasing in $\delta$, and nondecreasing in $t$.
\end{proposition}

The probability in \eqref{eqn:Random_circuits__Maximum} is a more stringent error measure than e.g.\ \eqref{eqn:Random_circuits__Distribution} is. The event $\{ \max_{0 \leq s \leq t} D_s<\delta\}$ implies after all that the error $ D_t $ has always been below the threshold $ \delta $ up to and including at time $ t $. The expected error $ \expectationWrt{D_t} {z_0} $ and distribution $ \probabilityWrt{ D_t > \delta }{z_0} $ only concern the error \emph{at} time $ t $. Additionally, \eqref{eqn:Random_circuits__Maximum} allows us to calculate the maximum number of gates that can be performed. That is,
$
\probabilityWrt{ \max_{0 \leq s \leq t} D_s > \delta }{z_0}
\leq \gamma
$
as long as
\begin{equation}
t 
\leq t^\star_{\delta,\gamma}
\triangleq \max \bigl\{ t \in \naturalNumbersZero \big| \probabilityWrt{ \max_{0 \leq s \leq t} D_s > \delta }{z_0} \leq \gamma \bigr\}.
\label{maximum_gates}
\end{equation}
In words: at most $t^\star_{\delta,\gamma}$ gates can be applied before an accumulated error of size at least $\delta$ occurred with probability at least $\gamma$. 

\smallskip\noindent
\emph{Proof of \eqref{eqn:Random_circuits__Distribution}.} It follows from \eqref{eqn:Definition_of_the_set_of_delta_bad_gate_pairs}, mutual exclusivity, and \eqref{eqn:Pzwt_for_homogeneous_Markov_chain} that
\begin{equation}
\probabilityWrt{ D_t > \delta }{z_0}
= \probabilityWrt{ Z_t \in \mathcal{B}_{\ket{\psi_0},\delta}^{\ket{\Psi_0}} }{z_0}
= \sum_{ w \in \mathcal{B}_{\ket{\psi_0},\delta}^{\ket{\Psi_0}} } \probabilityWrt{ Z_t = w }{z_0}
= \sum_{ w \in \mathcal{B}_{\ket{\psi_0},\delta}^{\ket{\Psi_0}} } ( P^t )_{z_0,w}.
\end{equation}
The right-hand side equals \eqref{eqn:Random_circuits__Distribution} in matrix notation. To obtain the expression for the expectation, directly apply the definition of expectation for a discrete random variable:
\begin{equation}
\expectationWrt{ D_t }{z_0}
= \sum_{ (x,y) \in \mathcal{G}_n^2 } \pnorm{ x \rho_0 x^\dagger - y \sigma_0 y^\dagger }{D} \probabilityWrt{ Z_t = (x,y) }{z_0}.
\label{eqn:Intermediate_calculation_of_the_expectation}
\end{equation}
Using \eqref{eqn:Pzwt_for_homogeneous_Markov_chain} and the definition of $d_{\ket{\psi_0}}^{\ket{\Psi_0}}$, this gives the result. 

\smallskip\noindent
\emph{Proof of \eqref{eqn:Random_circuits__Maximum}.} If $z_0 \in \mathcal{B}_{\ket{\psi_0},\delta}^{\ket{\Psi_0}}$, then $\probabilityWrt{ T_{\mathcal{B}_{\ket{\psi_0},\delta}^{\ket{\Psi_0}}} = 0 }{z_0} = 1$. If $z_0 \not\in \mathcal{B}_{\ket{\psi_0},\delta}^{\ket{\Psi_0}}$, then use \eqref{eqn:Definition_of_the_B_matrix_for_random_circuits} to write
\begin{align}\label{eq:probabilityhittingtime}
\probabilityWrt{ T_{\mathcal{B}_{\ket{\psi_0},\delta}^{\ket{\Psi_0}}} = s }{z_0}
\nonumber &
= \probabilityWrt{ 
Z_1 \not\in \mathcal{B}_{\ket{\psi_0},\delta}^{\ket{\Psi_0}}, 
\ldots, 
Z_{s-1} \not\in \mathcal{B}_{\ket{\psi_0},\delta}^{\ket{\Psi_0}}, 
Z_s \in \mathcal{B}_{\ket{\psi_0},\delta}^{\ket{\Psi_0}} 
}{z_0}
\nonumber \\ &
= \sum_{ z_1 \not\in \mathcal{B}_{\ket{\psi_0},\delta}^{\ket{\Psi_0}} } \cdots \sum_{ z_{s-1} \not\in \mathcal{B}_{\ket{\psi_0},\delta}^{\ket{\Psi_0}} } \smashoperator[r]{ \sum_{ z_s \in \mathcal{B}_{\ket{\psi_0},\delta}^{\ket{\Psi_0}} } } \probabilityWrt{ Z_1 = z_1, \ldots, Z_s = z_s }{z_0}
\nonumber \\ &
= e_{z_0}^\mathrm{T} \bigl( B_{\ket{\psi_0},\delta}^{\ket{\Psi_0}} \bigr)^{s-1} \bigl( P - B_{\ket{\psi_0},\delta}^{\ket{\Psi_0}} \bigr) \vectOnes{ \mathcal{B}_{\ket{\psi_0},\delta}^{\ket{\Psi_0}} }
\end{align}
in matrix notation. The result follows after summing \eqref{eq:probabilityhittingtime} for $ s=0,1,\ldots, t-1 $ by mutual exclusivity.

Note finally that for arbitrary $\delta_2 > \delta_1$, we have that $\mathcal{B}_{\ket{\psi_0}, \delta_2}^{\ket{\Psi_0}}\subseteq \mathcal{B}_{\ket{\psi_0}, \delta_1}^{\ket{\Psi_0}}$. As a consequence, 
\begin{equation}
\probabilityWrt{ T_{\mathcal{B}_{\ket{\psi_0}, \delta_2}^{\ket{\Psi_0}}} \leq t }{z_0} \leq \probabilityWrt{ T_{\mathcal{B}_{\ket{\psi_0}, \delta_1}^{\ket{\Psi_0}}} \leq t }{z_0}.
\end{equation}
This establishes that $\probabilityWrt{ T_{\mathcal{B}_{\ket{\psi_0}, \delta}^{\ket{\Psi_0}}} \leq t }{z_0}$ is nonincreasing in $\delta$. By positivity of the summands, $\probabilityWrt{ T_{\mathcal{B}_{\ket{\psi_0}, \delta}^{\ket{\Psi_0}}} \leq t }{z_0}$ is nondecreasing in $t$. \qed

\paragraph{Lower bound.} For general $\mathcal{B}_{\ket{\psi_0},\delta}^{\ket{\Psi_0}}$, the explicit calculation of \refEquation{eqn:Random_circuits__Maximum} can be numerically intensive. It is however possible to provide a lower bound of lower numerical complexity via the expected hitting time of the set $\mathcal{B}_{\ket{\psi_0},\delta}^{\ket{\Psi_0}}$. 

\begin{lemma}[Lower bound for random circuits] \label{lemma2}
For any set $\mathcal{A} \subseteq \mathcal{G}_n^2$, the expected hitting times of a homogeneous Markov chain are the solutions to the linear system of equations
$
\expectationWrt{T_\mathcal{A}}{z} = 0
$
for $z \in \mathcal{A}$, 
$
\expectationWrt{T_\mathcal{A}}{z} = 1 + \sum_{w \not\in \mathcal{A}} P_{z,w} \expectationWrt{T_\mathcal{A}}{w}
$ 
for $z \not\in \mathcal{A}$. Furthermore; for any $z_0 \in \mathcal{G}_n^2$, $\delta \geq 0$, $t = 0, 1, \ldots, \tau < \infty$:
\begin{equation}
\probabilityWrt{ \max_{ 0 \leq s \leq t } D_s > \delta }{z_0}
\geq 0 \vee \Bigl( 1 - \frac{  \expectationWrt{ T_{\mathcal{B}_{\ket{\psi_0},\delta}^{\ket{\Psi_0}} } }{z_0} }{ t + 1 } \Bigr).
\label{eqn:Lower_bound_on_the_maximum_probability_for_random_circuits}
\end{equation}
Here $ a\vee b\triangleq \max\{a, b\} $.
\end{lemma}

\noindent
\emph{Proof of \eqref{eqn:Lower_bound_on_the_maximum_probability_for_random_circuits}.} The first part is a standard result, see e.g.\ \cite[p.~202]{van2014performance}. The second part follows from Markov's inequality, i.e.,
\begin{equation}
\probabilityWrt{ \max_{ 0 \leq s \leq t } D_s \leq \delta }{z_0}
= \probabilityWrt{ T_{\mathcal{B}_{\ket{\psi_0},\delta}^{\ket{\Psi_0}}} > t }{z_0}
\leq \frac{ \expectationWrt{ T_{\mathcal{B}_{\ket{\psi_0},\delta}^{\ket{\Psi_0}}} }{z_0} }{ t + 1 }.
\end{equation}
That is it. \qed

As a consequence of \refLemma{lemma2},
$
    \probabilityWrt{ \max_{ 0 \leq s \leq t } D_s > \delta }{z_0} \geq \gamma
$
when
$
    t \geq { \expectationWrt{ T_{\mathcal{B}_{\ket{\psi_0},\delta}^{\ket{\Psi_0}}} }{z_0} } 
    \allowbreak
    / 
    ({ 1 - \gamma }) - 1,
$
and in particular
$
    \probabilityWrt{ \max_{ 0 \leq s \leq t } D_s > 0 }{z_0} > 0
$
when
$
    t 
    \geq 
    \expectationWrt{ T_{\mathcal{B}_{\ket{\psi_0},0}^{\ket{\Psi_0}}} }{z_0}.
$
The values in the right-hand sides are thus upper bounds to the number of gates $ t^\star_{\delta,\gamma}$ one can apply before $ \delta $ error has occurred with probability $ \gamma $:
\begin{equation}
t_{\delta,\gamma}^\star
\leq \expectationWrt{ T_{\mathcal{B}_{\ket{\psi_0},0}^{\ket{\Psi_0}}} }{z_0} \wedge \Bigl( \frac{ \expectationWrt{ T_{\mathcal{B}_{\ket{\psi_0},\delta}^{\ket{\Psi_0}}} }{z_0} }{ 1 - \gamma } - 1 \Bigr)
\end{equation}
for $\delta \geq 0, \gamma \in [0,1]$. Here, $a \wedge b \triangleq \min\{ a,b \}$.

\revisedPartBegin

\paragraph{Limitations of the method: types of quantum noise channels.}
The approach taken in this article is a hybrid between classical probability theory and quantum information theory. The results of this article are therefore not applicable to all quantum channels, and it is important that we signal you the limitations. 

As an illustrative example, consider the elementary circuit of depth $\tau = 1$ with $n=1$ qubit, in which the one gate is restricted to the Clifford group $\{ C_1, \ldots, C_{24} \}$, say. For such an elementary circuit, this article describes a classical stochastic process that chooses one of twenty-four quantum noise channel $\mathcal{F}^{(1)}, \ldots \mathcal{F}^{(24)}$ say according to some arbitrary classical probability distribution $\{ p_i(\rho) \}$, i.e.,
\begin{equation}
\rho_0
\to 
\rho_1 
= \mathcal{F}(\rho_0)
= 
\begin{cases}
\mathcal{F}^{(1)}(\rho_0) = C_1 \rho_0 C_1^\dagger & \textnormal{w.p. } p_1(\rho_0), \\
\mathcal{F}^{(2)}(\rho_0) = C_2 \rho_0 C_2^\dagger & \textnormal{w.p. } p_2(\rho_0), \\
\ldots & \\
\mathcal{F}^{(24)}(\rho_0) = C_{24} \rho_0 C_{24}^\dagger & \textnormal{w.p. } p_{24}(\rho_0). \\
\end{cases}
\label{eqn:Quantum_channel_F}
\end{equation}
Here, the classical probability distribution $\{ p_i(\rho) \}$ may be chosen arbitrarily, and depend on the initial quantum state $\rho_0$ as indicated. For this elementary quantum circuit of depth $\tau = 1$ with $n=1$ qubit, \eqref{eqn:Quantum_channel_F} characterizes the set of stochastic processes covered by our results in its entirety. 

For example, \refProposition{prop:Error_accumulation_distributions_for_random_circuits} cannot be applied to the deterministic process 
\begin{equation}
\rho_0 
\to \rho_1 
= 
\begin{cases}
\mathcal{E}^{(1)}(\rho_0) 
= (1-p) \rho_0 + p Y \rho_0 Y^\dagger
\enskip
\textnormal{w.p. } 1, \\
\end{cases}
\label{eqn:Quantum_channel_E1}
\end{equation}
nor to the deterministic process
\begin{equation}
\rho_0 
\to 
\rho_1 
= 
\begin{cases}
\mathcal{E}^{(2)}(\rho_0)
= (1-p) \rho_0 + \tfrac{p}{2} U \rho_0 U^\dagger + \tfrac{p}{2} U^\dagger \rho_0 U
\enskip
\textnormal{w.p. } 1. \\
\end{cases}
\label{eqn:Quantum_channel_E2}
\end{equation}
Here, $p \in (0,1)$ can be chosen arbitrarily and $U = \e{- \mathrm{i} \pi Y / 4 } $ is a Clifford gate. The reason is that
$
    \bigl( \mathcal{F}^{(1)} \neq \mathcal{F}^{(2)} \neq \cdots \neq \mathcal{F}^{(24)} \bigr)
    \neq 
    \bigl( \mathcal{E}^{(1)} = \mathcal{E}^{(2)} \bigr)
$
by the unitary freedom in the operator-sum representation \cite[Thm.~8.2]{nielsen_quantum_2011}. 
A meticulous reader will now note that the example quantum channels $\mathcal{E}^{(1)}$, $\mathcal{E}^{(2)}$ are however \emph{averages} of two particular stochastic processes $\mathcal{F}$. That is: if $p_I = 1 - p, p_Y = p$, then $\mathcal{E}^{(1)}(\rho) = \expectation{ \mathcal{F}(\rho) }$; or if $p_I = 1 - p, p_U = p_{U^\dagger} = \tfrac{p}{2}$, then $\mathcal{E}^{(2)}(\rho) = \expectation{ \mathcal{F}(\rho) }$.

An alternative way to understand what is going on, is to consider that we are describing the time-evolution of a density matrix, and that a density matrix expresses a subjective state of knowledge. The classical model described in this paper assumes that your best description of the system at each intermediate time step is a pure state, and this is not the case in quantum channels $\mathcal{E}^{(1)}$, $\mathcal{E}^{(2)}$. Your best description of the system at each intermediate time step is a pure state e.g.\ in randomized benchmarking when you are drawing classical random variables to randomly choose a quantum gate \emph{and} are being informed of their outcomes. Note finally that the expectation and probability operators in this paper are with respect to a classical stochastic process that drives a random choice of quantum gates, and that quantum measurements are thus not being modeled.

\paragraph{On how to construct the $P$ matrix.}
Both \refProposition{prop:Error_accumulation_distributions_for_random_circuits} and \refLemma{lemma2} rely on constructing the $P$ matrix. For illustrative purposes, we have written a script that will generate a valid $P$ matrix after a user inputs a vector describing (gate-dependent) error probabilities. The code is publicly available on TU/e's GitLab server at \url{https://gitlab.tue.nl/20061069/markov-chains-for-error-accumulation-in-quantum-circuits}. Additionally, we discuss an example in \refAppendixSection{appendix:example_to_calculate_the_lower_bound} for which we construct the $P$ matrix as well as evaluate the lower bound in \eqref{eqn:Random_circuits__Maximum}. 

\paragraph{An average over the trajectories of the Markov chain.} 
It is noteworthy that the results in \refProposition{prop:Error_accumulation_distributions_for_random_circuits} are averages over all noise trajectories that can be generated by the Markov chain. Consider e.g.\ \eqref{eqn:Random_circuits__Distribution}, which reads in matrix notation:
\begin{equation}
\probabilityWrt{ D_t > \delta }{z_0}
= e_{z_0}^T P^t 1_{ \mathcal{B}_{\ket{\psi_0},\delta}^{\ket{\Psi_0}} }
= \sum_{ w \in \mathcal{B}_{\ket{\psi_0},\delta}^{\ket{\Psi_0}} } ( P^t )_{z_0,w}.
\end{equation}
Expanding the matrix power, the right-hand side equals
\begin{align}
    &\underbrace{ \sum_{z_1 \in \mathcal{G}_n^2} \sum_{z_2 \in \mathcal{G}_n^2} \cdots \sum_{z_{t-1} \in \mathcal{G}_n^2} \sum_{ w \in \mathcal{B}_{\ket{\psi_0},\delta}^{\ket{\Psi_0}} } }_{ \textnormal{Term I} } 
    \underbrace{ P_{z_0,z_1} P_{z_1,z_2} \cdots P_{z_{t-1},w} }_{ \textnormal{Term II} }\\
   & = \expectationWrt{ \indicator{ Z_1 \in \mathcal{G}_n^2, \ldots, Z_{t-1} \in \mathcal{G}_n^2, Z_t \in \mathcal{B}_{\ket{\psi_0},\delta}^{\ket{\Psi_0}} } }{z_0}.
    \label{eqn:Sum_over_weighted_paths}
\end{align}
Here, Term I enumerates all possible length-$t$ trajectories of the Markov chain that start at some state $z_0 \in \mathcal{G}_n^2$ and end at any state $w \in \mathcal{B}_{\ket{\psi_0},\delta}^{\ket{\Psi_0}}$. Term II is the probability that the specific trajectory $z_0 \to z_1 \to z_2 \to \cdots \to z_{t-1} \to w$ occurs in this Markov chain. Consequently, \eqref{eqn:Sum_over_weighted_paths} is the expectation (average) of the random variable $\indicator{ Z_1 \in \mathcal{G}_n^2, \ldots, Z_{t-1} \in \mathcal{G}_n^2, Z_t \in \mathcal{B}_{\ket{\psi_0},\delta}^{\ket{\Psi_0}} }$ as indicated.
\revisedPartEnd

\subsubsection{The case of nonrandom circuits}

Suppose that the gate sequence $ \mathcal{U}_\tau= \lbrace U_{1},...,U_{\tau} \rbrace $ is fixed \emph{a priori} and that it is not generated randomly. Because the gate sequence is nonrandom, we have now that the faultless state $ \rho_{t}=X_t \rho_{0} {X^{\dagger}_t}$ is deterministic for times $t=0, 1, \ldots ,\tau $. On the other hand the potentially faulty state $ \sigma_{t}=Y_t \rho_{0} {Y^{\dagger}_t}$ is still (possibly) random. 

We can now use a lower dimensional Markov chain to represent the system. To be precise: we will now describe the process \revised{$\process{ Y_t }{t \geq 0}$ (and consequently $\process{ \sigma_t }{t \geq 0}$)} as an \emph{inhomogeneous Markov chain}. Its transition matrices will now be time-dependent and given element-wise by
\revised{
$
Q_{y,v}(t) 
= \probability{ Y_{t+1} = v | Y_t = y }
$
}
for $y, v \in \mathcal{G}_n, t \in \{ 0, 1, \ldots, \tau-1 \}$. Letting \revised{$Q^{(t)}_{y,v} \triangleq \probability{ Y_t = v | Y_0 = y }$} stand in for the probability that the process \revised{$\process{ Y_t }{ t \geq 0 }$} is at state \revised{$v$} at time $t$ starting from \revised{$y$}, we have by the Markov property \cite{bremaud2017discrete} that 
\begin{equation}
Q^{(t)}_{y,v}
= \bigl( \prod_{s=1}^t Q(s) \bigr)_{y, v}
\quad
\textrm{for}
\quad
y, v \in \mathcal{G}_n.
\label{eqn:Qzwt_for_inhomogeneous_Markov_chain}
\end{equation}
Note that the Markov chain modeled here is inhomogeneous, which is different from \refSection{sec:The_case_of_random_circuits}. In particular, the time-dependent transition matrix $Q(t)$ here cannot be expressed in terms of a power $P^t$ of a transition matrix $P$ on the same state space as in \refSection{sec:The_case_of_random_circuits}.

\emph{Example 3:} Consider the situation that the probability that an error occurs depends on which gate was applied last. If we assume that $\probability{ \Lambda_{t+1} = \lambda | Y_t = y} = \zeta_{y, U_{t+1}}(\lambda)$ are given distributions for $y \in \mathcal{G}_n$, $ t\in \{0, 1, \cdots, \tau-1\} $ on $\lambda \in \mathcal{G}_n$, we can alternatively write the elements of the transition matrices as
\begin{align}
    Q_{y, v}(t)
    &
    = \probability{ Y_{t+1} = v | Y_t = y}
    \nonumber \\ &
    = \sum_{ \lambda \in \mathcal{G}_n } \probability{ Y_{t+1} = v | Y_t = y, \Lambda_{t+1} = \lambda }\probability{ \Lambda_{t+1} = \lambda | Y_t = y}
    \nonumber \\ &
    = \sum_{ \lambda \in \mathcal{G}_n } \indicator{\lambda U_{t+1} y {\rho_0} {y}^\dagger U_{t+1}^\dagger{\lambda}^\dagger =v\rho_0 v^\dagger} \zeta_{y, U_{t+1}}(\lambda).
\end{align}
Here, we have used the law of total probability. 

\emph{Example 4:} If errors occur independently and with probability $\probability{ \Lambda_{t+1} = \lambda } = \zeta(\lambda)$, then \\$ Q_{y,v}(t)=\sum_{ \lambda \in \mathcal{G}_n } \indicator{ \lambda U_{t+1} y\rho_0 y^\dagger {U_{t+1}}^\dagger{\lambda}^\dagger = v\rho_0 v^\dagger } \zeta(\lambda). $

\paragraph{Results.} 

Now define the sets of $(\delta,t)$-bad gate pairs by
$
\mathcal{B}_{\ket{\psi_0},\delta}^{\ket{\Psi_0}, t}
\triangleq \bigl\{ x \in {\mathcal{U}_{n}}\big| \pnorm{\rho_t-  x \sigma_0 x^\dagger}{d} > \delta \bigr\}
$ for $\ket{\psi_0}, \ket{\Psi_0} \in \mathcal{S}^n$, $t \in \{ 0, 1, \allowbreak \ldots, \allowbreak \tau \}$, $\delta \geq 0$. Also define the matrices $B_{\ket{\psi_0}, \delta}^{\ket{\Psi_0}, t} \in [0,1]^{\cardinality{\mathcal{G}_n }\times\cardinality{\mathcal{G}_n }}$ element-wise by
\begin{equation}
\bigl( B_{\ket{\psi_0}, \delta}^{\ket{\Psi_0}, t}\bigr)_{y,v}
\triangleq
\begin{cases}
Q_{y,v}(t) & \textrm{if} \quad v \not\in \mathcal{B}_{\ket{\psi_0},\delta}^{\ket{\Psi_0}, t}, \\
0 & \textrm{otherwise},
\end{cases}
\label{eqn:Definition_of_the_B_matrix_for_fixed_circuits}
\end{equation}
for $t = 0, 1, \ldots, \tau $. Recall the notation introduced above \refProposition{prop:Error_accumulation_distributions_for_random_circuits}. Similarly enumerate in the vector ${d_{\rho_t}}$ the Schatten $d$-norms between any of the possibles states of $\sigma_t$ and the faultless state $\rho_{t}$. We state our second result:

\begin{proposition}[Error accumulation in nonrandom circuits]\label{pro:nonrandom circuits}
For any $ y_0\in\mathcal{G}_n $, $\delta \geq 0$, $t = 0, 1, \ldots, \tau < \infty$: the expected error is given by
$
	\expectationWrt{D_t} {y_0}
	=
	e_{y_0}^\mathrm{T} \bigl(\mathop \prod_{k = 1}^tQ(k)\bigr){d_{\rho_t}}.
$
Similarly, the distribution of error is given by
\begin{equation}
\probabilityWrt{ D_t > \delta }{y_0}
= e_{y_0}^\mathrm{T}\bigl(\mathop \prod \limits_{k = 1}^tQ(k)\bigr)\vectOnes{ \mathcal{B}_{\ket{\psi_0},\delta}^{\ket{\Psi_0}, t}}.
\label{eqn:Fixed_circuits__Distribution}
\end{equation}
Furthermore; if $y_0 \not\in \mathcal{B}_{\ket{\psi_0},\delta}^{\ket{\Psi_0}, 0}$, the distribution of maximum error is given by
\begin{equation}
\probabilityWrt{ \max_{0 \leq s \leq t} D_s > \delta }{y_0}
= \sum\limits_{s=0}^{t-1} \Bigl( e_{y_0}^\mathrm{T} \bigl( \mathop \prod \limits_{r = 0}^{s} B_{\ket{\psi_0},\delta}^{\ket{\Psi_0},r} \bigr) 
\times \bigl( Q(s+1) - B_{\ket{\psi_0},\delta}^{\ket{\Psi_0},s+1} \bigr) \vectOnes{ \mathcal{B}_{\ket{\psi_0},\delta}^{\ket{\Psi_0}, s+1} } \Bigr),
\label{eqn:Fixed_circuits__Maximum}
\end{equation}
and otherwise it equals one.
\end{proposition}

\smallskip\noindent
\emph{Proof of \eqref{eqn:Fixed_circuits__Distribution}.} From $\mathcal{B}_{\ket{\psi_0},\delta}^{\ket{\Psi_0}, t}$'s definition and mutual exclusivity it follows immediately that
\begin{equation}
\probabilityWrt{ D_t > \delta }{y_0}
= \probabilityWrt{ Y_t \in \mathcal{B}_{\ket{\psi_0},\delta}^{\ket{\Psi_0}, t} }{y_0}
= \sum_{  v \in \mathcal{B}_{\ket{\psi_0},\delta}^{\ket{\Psi_0}, t}} \probabilityWrt{ Y_t = v }{y_0} 
\label{eqn:trace_distance_delta}
\end{equation}
for $\ket{\psi_0}, \ket{\Psi_0}\in \mathcal{S}^n, \delta \geq 0 $. Using \eqref{eqn:Qzwt_for_inhomogeneous_Markov_chain} and continuing from \refEquation{eqn:trace_distance_delta}, we obtain
\begin{equation}
\probabilityWrt{ D_t > \delta }{y_0}
= \sum_{ v\in \mathcal{B}_{\ket{\psi_0},\delta}^{\ket{\Psi_0}, t}} e_{y_0}^\mathrm{T} \bigl(\mathop \prod \limits_{k = 1}^tQ(k) \bigr)_{y,v}.
\end{equation}
This simplifies to \eqref{eqn:Fixed_circuits__Distribution} in matrix notation. To obtain the expression for the expectation, apply the same arguments as were used for \refProposition{prop:Error_accumulation_distributions_for_random_circuits}, but use \eqref{eqn:Qzwt_for_inhomogeneous_Markov_chain} instead.

\smallskip\noindent
\emph{Proof of \eqref{eqn:Fixed_circuits__Maximum}.} We can again explicitly calculate the result using a hitting time analysis, but the expressions expand due to the time-dependency of $\mathcal{B}_{\ket{\psi_0},\delta}^{\ket{\Psi_0}, t}$. If $y_0 \in \mathcal{B}_{\ket{\psi_0},\delta}^{\ket{\Psi_0}, 0}$, then $\probabilityWrt{ \max_{ 0 \leq r \leq s } D_r > \delta }{y_0} = 1$. Otherwise
\begin{align}
&\probabilityWrt{ \lbrace\max_{ 0 \leq r \leq s-1 } D_r \leq \delta \rbrace \cap\lbrace D_s > \delta\rbrace}{y_0}  \\
&= \probabilityWrt{ 
Y_1 \not\in \mathcal{B}_{\ket{\psi_0},\delta}^{\ket{\Psi_0}, 1}, 
\ldots, 
Y_{s-1} \not\in \mathcal{B}_{\ket{\psi_0},\delta}^{\ket{\Psi_0}, s-1}, 
Y_s \in \mathcal{B}_{\ket{\psi_0},\delta}^{\ket{\Psi_0}, s} }{y_0}\\ 
&= \sum_{ y_1 \not\in \mathcal{B}_{\ket{\psi_0},\delta}^{\ket{\Psi_0}, 1} } \cdots  \sum_{ y_s \in \mathcal{B}_{\ket{\psi_0},\delta}^{\ket{\Psi_0}, s} }  \probabilityWrt{ Y_1 = y_1, \ldots, Y_s = y_s }{y_0}\label{eq:17} \\
&= \sum_{ y_1 \not\in \mathcal{B}_{\ket{\psi_0},\delta}^{\ket{\Psi_0},1} } \cdots \sum_{ y_{s-1} \in \mathcal{B}_{\ket{\psi_0},\delta}^{\ket{\Psi_0},s-1} } \sum_{ y_s \in \mathcal{B}_{\ket{\psi_0},\delta}^{\ket{\Psi_0},s} } \prod_{r=0}^{s-1} Q_{ y_r, y_{r+1}}(r).
\nonumber
\end{align}
Recalling \eqref{eqn:Definition_of_the_B_matrix_for_fixed_circuits}, we can equivalently write \eqref{eq:17} in matrix notation as
\begin{equation}
\probabilityWrt{ \lbrace\max_{ 0 \leq r \leq s-1 } D_r \leq \delta \rbrace \cap\lbrace D_s > \delta\rbrace}{y_0}
= e_{y_0}^\mathrm{T} \bigl( \prod_{r = 1}^{s-1} B_{\ket{\psi_0}, \delta}^{\ket{\Psi_0},r} \bigr) \bigl( Q(s) - B_{\ket{\psi_0}, \delta}^{\ket{\Psi_0},s} \bigr) \vectOnes{ \mathcal{B}_{\ket{\psi_0},\delta}^{\ket{\Psi_0}, s} }. 
\label{eqn:Precise_hitting_time_distribution_for_fixed_circuits}
\end{equation}
Summing \eqref{eqn:Precise_hitting_time_distribution_for_fixed_circuits} over $s = 0, 1, \ldots, t - 1$ completes the proof by mutual exclusivity. \qed

\revisedPartBegin
\paragraph{On how to construct the $Q$ matrix.}
The script that we created that can generate example $P$ matrices, can also generate valid $Q$ matrices after the user inputs a vector describing (gate-dependent) error probabilities. Recall that this code is available on TU/e's GitLab server here: \url{https://gitlab.tue.nl/20061069/markov-chains-for-error-accumulation-in-quantum-circuits}.
\revisedPartEnd

\subsubsection{State space reduction in stabilizer circuits}

The set of stabilizer gates~\cite{gottesman1997stabilizer} for a state $ \ket{\psi}$ is defined as the set of gates $ \mathcal{M} \in \mathcal{G}_{n}\setminus I^{\otimes n}$ that satisfy $ \mathcal{M}\ket{\psi}=e^{i\gamma}\ket{\psi}$ for some $ \gamma\in\mathbb{R} $. Since $e^{i\gamma}$ is a global phase that cannot be observed, $ \mathcal{M}\ket{\psi}=e^{i\gamma}\ket{\psi}$ can also be understood as part of an equivalence class $ \mathcal{M}\ket{\psi}\equiv\ket{\psi}$. The state $ \ket{\psi} $ in $ \mathcal{M}\ket{\psi}\equiv\ket{\psi}$ is called the \emph{stabilizer state}~\cite{garcia2017geometry}. For one qubit and in case of the Pauli group, examples include $ \ket{0} $, $ \ket{1} $, and $ \ket{\pm}=(1/2)(\ket{0}\pm\ket{1}) $. Remark~\ref{claim:2} shows that there exist $ 2^{n} $ stabilizer states for any gate $ \mathcal{M}\in \mathcal{G}_{n}\setminus I^{\otimes n} $. Its proof is relegated to \refAppendixSection{appendix:Number_of_stabilizer_states}.

\begin{remark}\label{claim:2}
For any gate $ \mathcal{M}\in \mathcal{G}_{n}\setminus I^{\otimes n} $ there are $ 2^{n} $ states $ \ket{\psi_{0}} $ that satisfy $ \mathcal{M}\ket{\psi_{0}}=e^{i\gamma}\ket{\psi_{0}}$ for some $\gamma \in \realNumbers$.
\end{remark}

The advantage of starting a quantum circuit from a stabilizer state is that the state space is smaller. It moreover can be proved that, under the assumptions of \refSection{sec:Model}, when starting initially from a stabilizer state, all states reached during the quantum computation will themselves be stabilizer states. Define the set of \emph{reachable density matrices} from an initial state $ \ket{\psi_{0}} \in \mathcal{S}^n$, by
\begin{equation}\label{reachable_state}
\mathcal{R}_{ \ket{\psi_0}} 
\triangleq \bigl\{ g \ket{ \psi_0 } \big| g \in  \mathcal{G}_{n} \bigr\}.
\end{equation} 
The exact number of reachable states can be calculated by the method in \refAppendixSection{appendix1}. Taking the Clifford group gates on two qubits as an example, the number of gates $ |\mathcal{C}_2|=11520 $. However, there are just $ 60 $ reachable states if the initial state is $ \ket{00} $. The proof of Remark~\ref{Lemma:5} can be found in \refAppendixSection{appendix:A_stabilizer_state_follows_after_a_stabilizer_state}.

\begin{remark}\label{Lemma:5}
Given a gate $\mathcal{M}\in\mathcal{G}_{n}\setminus I^{\otimes n}$ and a state $ \ket{\psi_{0}}\in\mathcal{S}_{n} $ such that $ \mathcal{M}\ket{\psi_{0}}=e^{i\gamma}\ket{\psi_{0}}$ for some $ \gamma\in\mathbb{R} $, then for any state $ \ket{\psi_{1}}\in{\mathcal{R}_{\ket{\psi_{0}}}} $ there exists an $ \mathcal{H}\in \mathcal{G}_n \setminus I^{\otimes n}$ such that $ \mathcal{H}\ket{\psi_{1}}=e^{i\gamma}\ket{\psi_{1}}$.
\end{remark}

A consequence of Remark~\ref{Lemma:5} is namely that for any reachable state $ \ket{\Psi} $ there are at least two different gates $\mathcal{M}_i, \mathcal{M}_j\in\mathcal{G}_{n}$ whose corresponding states $\mathcal{M}_i\ket{\psi_{0}}$ and $\mathcal{M}_j\ket{\psi_{0}}$ are equivalent (up to a phase) to same state $ \ket{\Psi} $, since $\mathcal{M}_i\ket{\psi_{0}}\equiv\mathcal{M}_j\ket{\psi_{0}}\equiv  \ket{\Psi}$ if we let $ \ket{\Psi}=\mathcal{M}_i\ket{\psi_{0}} $ and $ \mathcal{M}_j=\mathcal{H}\mathcal{M}_i $. The number of reachable states $|\mathcal{R}_{ \ket{\psi_0}}|$ is thus upper bounded by $1/2|\mathcal{G}_n |$ when starting from a stabilizer state.

\subsection{Continuous, random error accumulation (one-qubit case)}
\label{sec:Continuous_random_error_accumulation_one_qubit_case}

In this section, we analyze the case where a single qubit:
\begin{enumerate}
\item receives a random perturbation on the Bloch sphere after each $s$-th unitary gate according to a continuous distribution $p_s(\alpha)$, and
\item depolarizes to the completely depolarized state $I/2$ with probability $q\in[0, 1]$ after each unitary gate,
\end{enumerate}
by considering it an absorbing random walk on the Bloch sphere. The key point leveraged here is that the trace distance is invariant under rotations. Hence a sufficiently symmetric random walk distribution will give the error probabilities.

\paragraph{Model.} Let $R_0$ be an initial point on the Bloch sphere. Every time a unitary quantum gate is applied, the qubit is rotated and receives a small perturbation. This results in a random walk ${\left\lbrace  R_t \right\rbrace }_{t \geq 0}$ on the Bloch sphere for as long as the qubit has not depolarized. Because the trace distance is invariant under rotations and since the rotations are applied both to $ \rho_t $ and $ \sigma_t $, we can ignore the rotations. We let $\nu$ denote the random time at which the qubit depolarizes. With the usual independence assumptions, $\nu \sim \mathrm{Geometric}(q)$.

Define $\mu_t(\vect{r})$ for $t < \nu$ as the probability that the random walk is in a solid angle $\Omega$ about $r$ (in spherical coordinates) conditional on the qubit not having depolarized yet. That is,
\begin{equation}
\probability{ R_t \in \mathcal{S} | \nu > t }
\triangleq\int_{ \mathcal{S} } \mu_t(r) \text{d}{\Omega(r)}.
\end{equation}
We assume without loss of generality that $R_0 = \hat{z}$. From~\cite{roberts1960random}, the initial distribution is then given by
\begin{equation}
\mu_0
= \sum_{n=0}^\infty \frac{2n+1}{4\pi} P_n( \cos{\theta} ).
\end{equation}
Here, the $P_n(\cdot)$ denote the Legendre polynomials. Also introduce the shorthand notation
\begin{equation}
    \Lambda_{n,t} 
    \triangleq \prod_{s=1}^t \int_0^\pi P_n( \text{cos}{\alpha} ) \text{d}{p_s}(\alpha)
\end{equation}
for convenience. As we will see in \refProposition{prop:Error_accumulation_in_a_single_qubit} in a moment, these constants will turn out to be the coefficients of an expansion for the expected trace distance (see \eqref{eqn:Single_qubit__Expectation}).
Recall that here, $p_s(\alpha)$ denotes the probability measure of the angular distance for the random walk on the Bloch sphere at time $ t $ (see (i) above). In particular: if $p_t(\alpha) = \delta(\alpha)$ for all $t \geq 0$ meaning that each step is taken into a random direction but exactly of angular length $\alpha$, then $\Lambda_{n,t} = ( P_n( \cos{\alpha} ) )^t$. From \cite{roberts1960random}, it follows that after $t$ unitary quantum gates have been applied without depolarization having occurred,
\begin{equation}
\mu_t
= \sum_{n=0}^\infty \frac{2n+1}{4\pi} \Lambda_{n,t} P_n( \cos{\theta} ).
\label{eqn:rhot}
\end{equation}

\paragraph{Results.} In this section we specify $ D_t $ as the trace distance. We are now in position to state our findings:

\begin{proposition}[Single qubit]
\label{prop:Error_accumulation_in_a_single_qubit}
For $0\leq\delta\leq 1$, $t\in\naturalNumbersPlus$: the expected trace distance satisfies
\begin{equation}
    \expectation{D_t} 
    = \tfrac{1}{2} - (1-q)^t \Bigl( \tfrac{1}{2} + 2 \sum\limits_{n=0}^\infty \frac{ \Lambda_{n,t} }{ (2n-1) (2n+3) } \Bigr).
    \label{eqn:Single_qubit__Expectation}
\end{equation}
The distribution of the trace distance is given by 
\begin{align}
\probability{ D_t \leq \delta } &= \indicator{ \tfrac{1}{2} \in [0,\delta] } \bigl( 1 - (1-q)^t \bigr) \\
&+
(1-q)^t \sum_{n=0}^\infty (2n+1) \Lambda_{n,t} \sum_{r=1}^{n+1} (-1)^{r+1} \delta^{2r} C_{r-1} \binom{ n+r-1 }{ 2(r-1) }.
\label{eqn:Single_qubit__Distribution}
\end{align}
Here, the $C_{r}$ denote the Catalan numbers.
Alternative forms include:
\begin{gather}
    \probability{ D_t \leq \delta | \nu > t } = \delta^2 \sum_{n=0}^\infty (2n+1) \Lambda_{n,t} {_2}F_1( -n, n+1, 2; \delta^2 )
    ,
    \quad
    \textnormal{and} 
    \\ 
    \probability{ D_t \leq \delta | \nu > t } = \delta^2 \sum_{n=0}^\infty (2n+1) \Lambda_{n,t} \frac{n!}{(2)_n} P_n^{(1,-1)}( 1 - 2 \delta^2 )
\end{gather}
with ${_2}F_1(a,b,c;z)$ the Hypergeometric function, $(\cdot)_n$ the Pochhammer symbol, and $P_n^{(\alpha,\beta)}(x)$ the Jacobi polynomials.
Finally; the distribution of maximum trace distance is lower bounded by
\begin{equation}
\probability{ \max_{0 \leq s \leq t} D_s \leq \delta | \nu > t } \geq
 0 \vee \Bigl(1 - t + \delta^2 \sum_{s=1}^t \sum_{n=0}^\infty (2n+1) \Lambda_{n,s} \frac{n!}{(2)_n} P_n^{(1,-1)}( 1 - 2 \delta^2 )\Bigr).
\label{eqn:Single_qubit__Maximum}
\end{equation}
\end{proposition}

\smallskip\noindent
\emph{Proof of \eqref{eqn:Single_qubit__Expectation}.} By the law of total expectation, we have
\begin{equation*}
\expectation{D_t} 
= \expectation{ D_t | \nu > t } \probability{ \nu > t } + \expectation{ D_t | \nu \leq t } \probability{ \nu \leq t }.
\end{equation*}
Since $\nu \sim \mathrm{Geometric}(q)$, we have that
\begin{equation*}
\probability{ \nu > t } = 1 - \probability{ \nu \leq t } = (1-q)^t.
\end{equation*}
Note additionally that $D_t = 1/2$ whenever $t \geq \nu$. Therefore
\begin{equation*}
\expectation{D_t} 
= \expectation{ D_t | \nu > t } (1-q)^t + \tfrac{1}{2} \bigl( 1 - (1-q)^t \bigr)
= \tfrac{1}{2} + \bigl( \expectation{ D_t | t < \nu } - \tfrac{1}{2} \bigr) (1-q)^t .
\end{equation*}

We now calculate $\expectation{ D_t | \nu > t }$ using \eqref{eqn:rhot} and the Bloch sphere representation:
\begin{align}
\expectation{ D_t | \nu > t }
&= \sum_{n=0}^\infty \frac{2n+1}{4\pi} \Lambda_{n,t} \int_0^\pi 2 \pi \sin{\theta} \sin{ \frac{\theta}{2} }  P_n( \cos\theta ) \text{d}{\theta}\\
&= \sum_{n=0}^\infty \frac{2n+1}{2} \Lambda_{n,t}\int_{-1}^{1}\sqrt{\frac{1-x}{2}} P_n(x) \text{d}{x}. \label{equation15}
\end{align}
Also recall two facts about the Legendre polynomials: the recurrence relation in~\cite{grosjean1985theory} states that
\begin{equation}\label{equation16}
P_n(x)=\frac{1}{2n+1}\left(P'_{n+1}(x)-P'_{n-1}(x)\right),
\end{equation}
and Rodrigues formula~\cite[(8.6.18)]{abramowitz1965handbook} states that
\begin{equation}\label{equation19}
P_n(x)=\dfrac{1}{2^{n}n!}\dfrac{\d{}^{n}}{\d{x^{n}}}{(x^2-1)}^{n}.
\end{equation}
Using \eqref{equation16}, \eqref{equation19}, and integration by parts, we then obtain
\begin{equation}\label{equation20}
\int_{-1}^{1}\sqrt{\frac{1-x}{2}} P_n(x) \text{d}{x}
= \frac{1}{2n+1} \Bigl(-\int_{-1}^1 \frac{P_{n+1}(x)}{2\sqrt{2-2x}} \d{x}+\int_{-1}^1 \frac{P_{n-1}(x)}{2\sqrt{2-2x}} \d{x} \Bigr).
\end{equation}
We have by \cite[(12.4)]{arfken1999mathematical} that the generating function of the Legendre polynomials is given by
\begin{equation}\label{equation21}
\sum\limits_{m=0}^\infty P_m(x)s^m=\frac{1}{\sqrt{1-2xs+s^2}}.
\end{equation}
Based on \eqref{equation21} with $ t=1 $ and the orthogonality of Legendre polynomials,
\begin{equation}
\int_{-1}^1 \frac{P_{n}(x)}{\sqrt{2-2x}} \d{x}
= \int_{-1}^1{P_{n}(x)}\sum\limits_{m=0}^\infty P_m(x) \d{x} 
= \sum\limits_{m=0}^\infty\int_{-1}^1{P_{n}(x)}{P_{m}(x)} \d{x}
= \frac{2}{2n+1}. \label{equation22}
\end{equation}
Here, we have used Lebesgue's dominated convergence theorem with $ | P_{n}(x) |\leq 1$ $ \forall n $.
Therefore, continuing from \eqref{equation15} using \eqref{equation20} and \eqref{equation22}, 
\begin{align}
\expectation{ D_t | \nu > t } 
&
= \sum_{n=0}^\infty \frac{2n+1}{2} \Lambda_{n,t} \Bigl( -\int_{-1}^1 \frac{P_{n+1}(x)}{2\sqrt{2-2x}} \d{x} + \int_{-1}^1 \frac{P_{n-1}(x)}{2\sqrt{2-2x}} \d{x} \Bigr)
\nonumber \\ & 
= \sum_{n=0}^\infty \frac{2n+1}{2} \Lambda_{n,t}\frac{-4}{(2n-1)(2n+1)(2n+3)}. 
\end{align}
Simplifying gives the result.

\smallskip\noindent
\emph{Proof of \eqref{eqn:Single_qubit__Distribution}.} Similar to above we have by the law of total probability that
\begin{equation*}
\probability{ a \leq D_t \leq b }
= \probability{ a \leq D_t \leq b | \nu \leq t } \probability{ \nu \leq t }
 + \probability{ a \leq D_t \leq b | \nu > t } \probability{ \nu > t },
\end{equation*}
and we note now that $\probability{ a \leq D_t \leq b | \nu \leq t } = \indicator{ \tfrac{1}{2} \in [a,b] }$. Therefore
\begin{equation}
\probability{ a \leq D_t \leq b }
= \indicator{ \tfrac{1}{2} \in [a,b] } \bigl( 1 - (1-q)^t \bigr) 
 + \probability{ a \leq D_t \leq b | \nu > t } (1-q)^t.
\end{equation}

We now calculate $\probability{ a \leq D_t \leq b | \nu > t }$; again using \eqref{eqn:rhot}. Let $0 \leq a \leq b \leq 1$. From the equivalence of the events
\begin{equation*}
\bigl\{ a \leq D_t \leq b \bigr\}
= \bigl\{ 2 \arcsin(a) \leq \Theta_t \leq 2 \arcsin(b) \bigr\},
\end{equation*}
where $\Theta_t$ denotes the polar angle of $R_t$, it follows that 
\begin{align}
\mathbb{P}\left[ a \leq D_t \leq b \right]& = \bigl( 1 - (1-q)^t \bigr) \indicator{ \tfrac{1}{2} \in [a,b] }\\
&+ (1-q)^t \sum_{n=0}^\infty \frac{2n+1}{4\pi} \Lambda_{n,t} \int_{ 2 \arcsin{a} }^{ 2 \arcsin{b} } 2 \pi \sin{\theta} P_n( \cos{\theta} ) \text{d}{\theta}.
\label{eq:padtb}
\end{align}
Now let $0\leq\delta \leq 1$. Continuing from \eqref{eq:padtb}, since $ \cos(2\arcsin\delta)=1-2\delta^2 $ for $ \delta\in [0, 1] $ and letting $ \cos\theta=x $,
\begin{align}
\mathbb{P}\left[ D_t \leq \delta | \nu > t \right]
&= \sum_{n=0}^\infty \frac{2n+1}{4\pi} \Lambda_{n,t} \int_{0}^{ 2 \arcsin{\delta} } 2 \pi \sin{\theta} P_n( \cos{\theta} ) \text{d}{\theta}\\
&=\sum_{n=0}^\infty \frac{2n+1}{2} \Lambda_{n,t} \int_{1-2\delta^2}^{1}P_n(x) \text{d}{x}.
\end{align}
By the explicit representation of Rodrigues' formula~\cite[(8.6.18)]{abramowitz1965handbook},
\begin{align}
\mathbb{P}\left[ D_t \leq \delta | \nu > t \right]
&= \sum_{n=0}^\infty \frac{2n+1}{2} \Lambda_{n,t} \int_{1-2\delta^2}^{1} \sum_{k=0}^n \binom{n}{k}\binom{n+k}{k}\Bigl(\dfrac{x-1}{2}\Bigr)^k\text{d}{x}\\
&= \sum_{n=0}^\infty{(2n+1)} \Lambda_{n,t} \sum_{k=0}^n \binom{n}{k}\binom{n+k}{k}\dfrac{(-1)^{k}}{k+1}\delta^{2(k+1)}.
\end{align}
Finally, let $ r=k+1 $, such that
\begin{align}
\mathbb{P}\left[ D_t \leq \delta | \nu > t \right]
&= \sum_{n=0}^\infty {(2n+1)} \Lambda_{n,t} \sum_{r=1}^{n+1} \binom{n}{r-1}\binom{n+r-1}{r-1}\dfrac{(-1)^{r-1}}{r}\delta^{2r}\\
&= \sum_{n=0}^\infty {(2n+1)}\Lambda_{n,t} \sum_{r=1}^{n+1} (-1)^{r-1} \delta^{2r} C_{r-1} \binom{ n+r-1 }{ 2(r-1) }.
\nonumber 
\end{align}

\smallskip\noindent
\emph{Proof of \eqref{eqn:Single_qubit__Maximum}.} This follows directly after applying De Morgan's law and Boole's inequality, i.e.,
\begin{align}
&
\mathbb{P}[ \max_{0 \leq s \leq t} D_s \leq \delta | \nu > t ]
= \mathbb{P}\Bigl[ \bigcap_{s=0}^t \Bigl\{ D_s \leq \delta\Bigr\}\Big| \nu > t\Bigr]\\
&= \mathbb{P}\Bigl[ \Bigl(\bigcup_{s=0}^t \bigl\{ D_s > \delta \bigr\}\Bigr)^{c} \Big| \nu > t\Bigr]
\nonumber=1-\mathbb{P}\Bigl[\bigcup_{s=0}^t \bigl\{ D_s >\delta \bigr\} \Big| \nu > t\Bigr]
\\ &\geq 1 - \sum_{s=0}^t \mathbb{P}[ D_s > \delta | \nu > t]
=1-t+\sum_{s=0}^t \mathbb{P}[ D_s \leq \delta | \nu > t].
\end{align}
That is it. \qed
\section{Simulations}
\label{sec:Numerical_results}

\revisedPartBegin
In this section, we investigate and validate our results numerically. This section also serves to illustrate the models. We also compare our results to the following traditional error calculation and fitting method. 

\paragraph{Fit method using just a depolarizing channel} First, one readily calculates the expected trace distance of a depolarizing quantum channel \cite[p.~378]{nielsen_quantum_2011} 
\begin{equation}
\rho_0 
\to \rho_1 = \mathcal{E}(\rho_0)
= \dfrac{\mu}{2} I + (1-\mu) \rho_0
\quad
\textnormal{w.p. } 1
\label{equ:depolarizing_channel}
\end{equation}
when repeated $t \in \naturalNumbersPlus$ times as a function of its decay parameter $\mu \in [0,1]$. To see how, note that after $t$ applications of this depolarizing channel, the quantum state would be $ \mathcal{E}^{t}(\rho) = \tfrac{1}{2} ( 1 - (1-\mu)^t ) I + (1-\mu)^t \rho$ w.p.\ one. The trace distance after $t$ depolarizing channels is thus
\begin{equation}
D_t 
= \tfrac{1}{2} ( 1 - (1-\mu)^t ) 
\quad \textnormal{w.p. } 1.
\label{equ:error_depolarizing_channel}
\end{equation}
Next, one fits \eqref{equ:error_depolarizing_channel} to experimental or numerical data using e.g.\ the method of least squares. This curve follows the data as well as it can (but not necessarily perfect), and the corresponding fit parameter $\mu^{\mathrm{fit}}$ is returned. 

It is insightful to consider the difference between \eqref{equ:depolarizing_channel}, \eqref{equ:error_depolarizing_channel} and the result in \refProposition{prop:Error_accumulation_in_a_single_qubit}. \refProposition{prop:Error_accumulation_in_a_single_qubit} namely models a different type of error channel, specifically one in which the qubit can depolarize at each step according to a classical probability $\mu \in [0,1]$. Substituting $\Lambda_{n,t} = 0$ for all $n,t$ so that the random perturbations of the model in \refSection{sec:Continuous_random_error_accumulation_one_qubit_case} are neglected and only depolarization is included, this model tells us that $\mathcal{E}^t(\rho)$ equals either $\rho$ w.p.\ $(1-\mu)^t$ or $I/2$ w.p.\ $1-(1-\mu)^t$. Consequently, under this model,
\begin{equation}
D_t 
= \begin{cases}
0 & \textnormal{w.p. } (1-\mu)^t \\
\tfrac{1}{2} & \textnormal{w.p. } 1 - (1-\mu)^t, \\
\end{cases}
\quad
\expectation{D_t} 
= \tfrac{1}{2} ( 1 - (1-\mu)^t ).
\end{equation}
Note that the expectation here equals \eqref{equ:error_depolarizing_channel} by chance. If $\Lambda_{n,t} \neq 0$, this would not have been the case.
\revisedPartEnd

\subsection{Error accumulation in randomized benchmarking}
\label{sec:Error_accumulation_in_randomized_benchmarking}

We \revised{will first} consider error accumulation in single-qubit randomized benchmarking. In each randomized benchmarking simulation experiment, the initial state is set to $ \ket{1} $ and subsequently $ \tau-1 $ gates are selected one by one from the Clifford group $ \mathcal{C}_1 $ uniformly at random. Finally, based on the experimental setup in \cite{xia2015randomized}, we add a $ \tau $-th gate that transfers the state to $ \ket{0} $ in the absence of errors. For simplicity we specify $d = 1$ and thus discuss the trace distance throughout this section.

\subsubsection{Pauli and Clifford channel errors}

\begin{figure}[!hbtp]
\begin{center}
\includegraphics[
width=0.7\linewidth
]%
{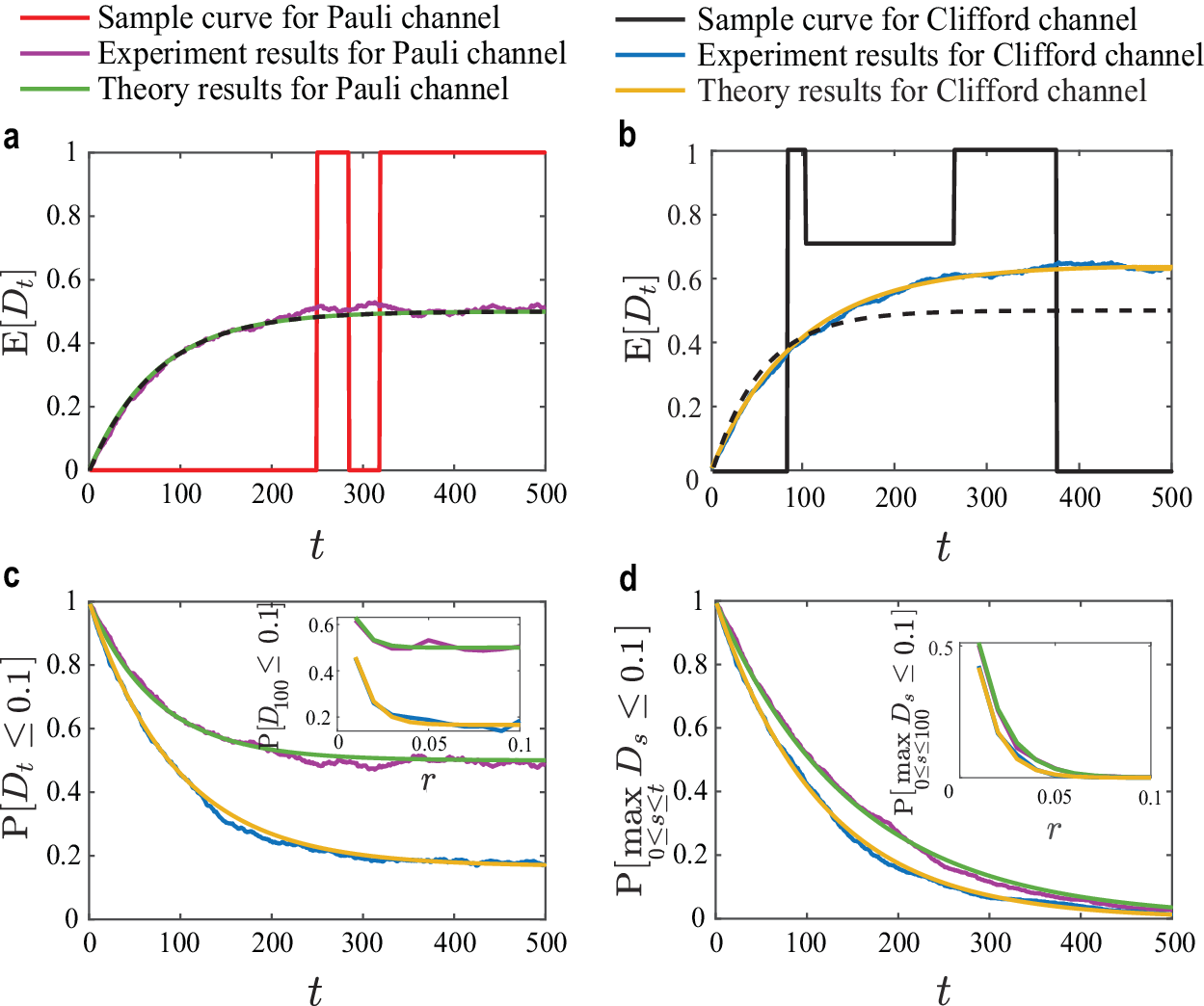}%
\caption{The error accumulation based on Pauli and Clifford channels in randomized benchmarking. Two error thresholds $ \delta $ are considered, $ \delta=1/10 $ (figures a, c, d) and $ \delta=1/5 $ (figure b). The simulation results are calculated from $ 1000 $ independent randomized benchmarking experiments. \revised{The dashed, black curves are fits of \eqref{equ:error_depolarizing_channel} to the sample averages of the numerical data. The resulting fit parameters are $ \mu^{\mathrm{fit}} \approx 0.013 $ (figure a) and $ \mu^{\mathrm{fit}} \approx 0.018 $ (figure b), respectively. }} 
\label{fig_randomized_benchmaring}%
\end{center}
\end{figure}

We consider two kinds of error models: Pauli channels and Clifford channels. For the Pauli channel model, let the probability that no noise occurs be $ \mathbb{P}(\Lambda=I)=1-r$, and the probabilities of every noise type occurring be $ \mathbb{P}(\Lambda=X)= \mathbb{P}(\Lambda=Y)= \mathbb{P}(\Lambda=Z)=r/3 $, where $ r\in [0, 1] $.  For the Clifford channel model, let the probability of no noise occurring be $ \mathbb{P}(\Lambda=I)=1-r$, and the probabilities of every other gate type occurring equal $ r/23 $. In \refFigure{fig_randomized_benchmaring} the parameter $ r $ is set to $ 1/100 $. Two error thresholds $ \delta $ are considered: $ \delta=1/10 $ (a, c, and d) and $ \delta=1/5 $ (b). The insets show the influence of parameter $ r $ on the probability of error in \eqref{eqn:Random_circuits__Distribution} and the probability of maximum error in \eqref{eqn:Random_circuits__Maximum} at time $ t=100 $. The results in \refFigure{fig_randomized_benchmaring} illustrate the theoretical results for the probability of error \eqref{eqn:Random_circuits__Distribution}, the expectation of the trace distance, and the probability of maximum error \eqref{eqn:Random_circuits__Maximum}, and their validity is supported by these simulations. \refFigure{fig_randomized_benchmaring} also illustrates that different error models lead to different error accumulation behaviors. \revised{The two sample curves in \refFigure{fig_randomized_benchmaring}a and \refFigure{fig_randomized_benchmaring}b (the solidly drawn step functions) show the trace distance $ D_t $ between the faultless state $ \rho_t $ and the faulty state $ \sigma_t $ in two independent randomized benchmarking experiments. The dashed lines indicate our fits of \eqref{equ:error_depolarizing_channel} to the sample average of the numerical data. Note that the numerical sample average of the trace distance in \refFigure{fig_randomized_benchmaring}a can be fitted perfectly -- this is because under the present assumptions, the trace distance here is in fact geometrically distributed. The case depicted in \refFigure{fig_randomized_benchmaring}b is however different and does not satisfy a simple geometric distribution, and we can see that the traditional fit method disagrees in the limit. This is because we are dealing with two different error models.}

\subsubsection{Influence of the initial state on Pauli error accumulation}\label{section31}

In this section we consider the influence of the initial state on Pauli error accumulation. We ignore the last gate of randomized benchmarking for simplicity. Each gate is selected one by one from the Pauli group uniformly at random. The error model described above is considered again and the parameter $ r $ is set to $ 1/5 $.

\begin{figure}[!hbtp]
\begin{center}
\includegraphics[
width=0.6\linewidth
]%
{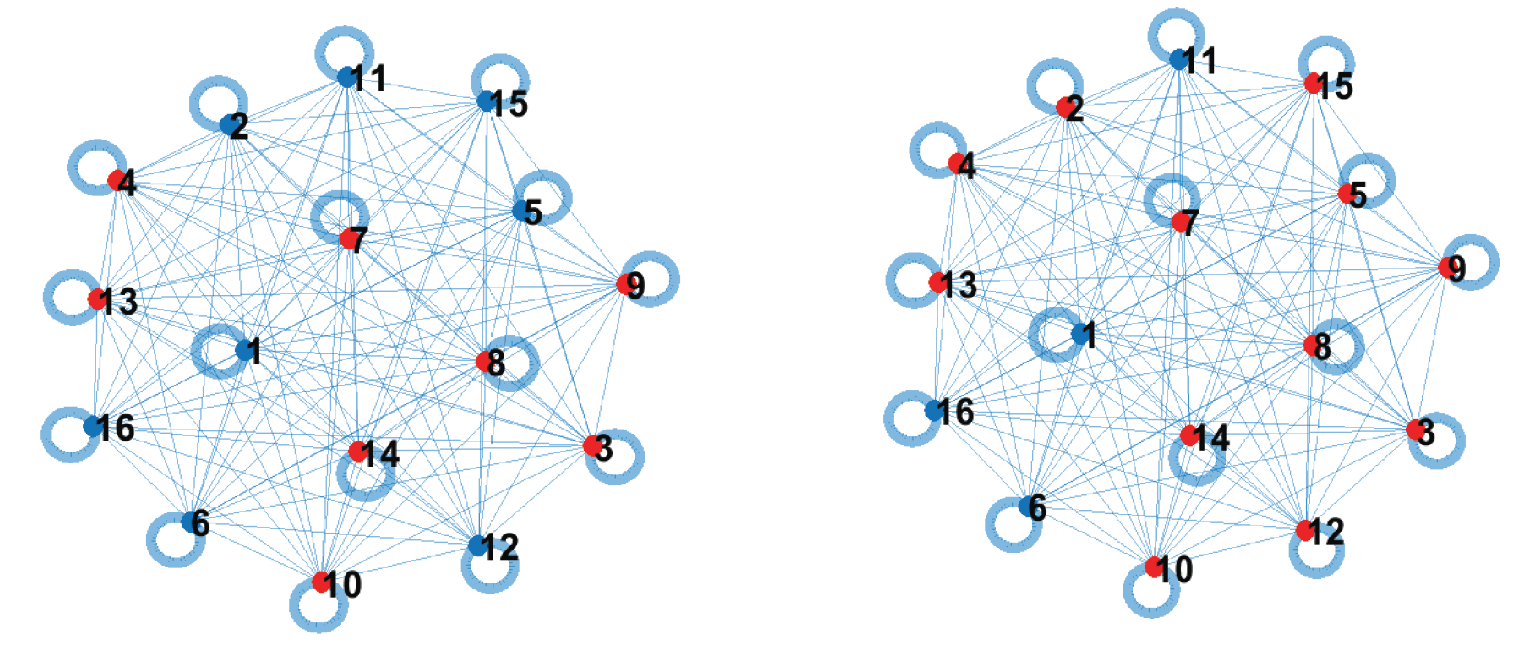}%
\caption{State transition diagram for different initial states: $\ket{\zeta_0}$ (left) and $\ket{\xi_{0}}$ (right), and the error threshold $\delta=1/5 $. The red nodes show the bad state pairs in $\mathcal{B}_{\ket{\zeta_0},\delta}^{\ket{\zeta_0}}$ and $\mathcal{B}_{\ket{\xi_0},\delta}^{\ket{\xi_0}}$, respectively, in which the trace distances are larger than $ \delta $.}
\label{fig_pauli_1a}%
\end{center}
\end{figure}

\begin{figure}[!hbtp]
\begin{center}
\includegraphics[
width=0.7\linewidth
]%
{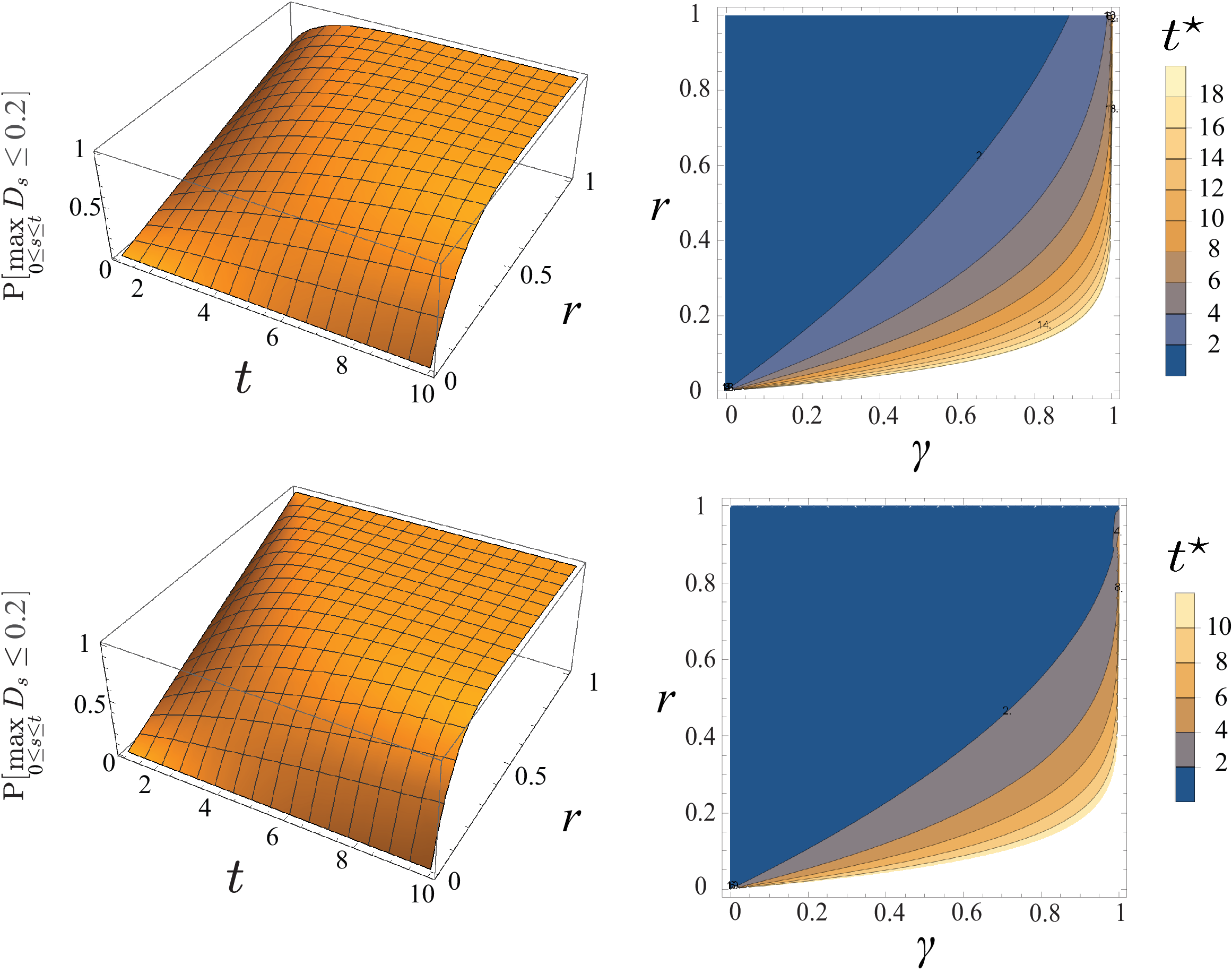}%
\caption{Pauli channel error accumulation on single-qubit randomized benchmarking when starting from different initial states: $\ket{\zeta_0}$ (top) and $\ket{\xi_{0}}$ (bottom). The error threshold is set to $\delta=1/5 $.}
\label{fig_pauli_1}%
\end{center}
\end{figure}

\refFigure{fig_pauli_1a} shows the state transition diagram for two different initial states: $\ket{\zeta_0}= \sqrt{7/10}\ket{0}+ \sqrt{3/10}\ket{1}$ and $\ket{\xi_{0}}= \sqrt{4/5}\ket{0}+ \sqrt{1/5}\ket{1}$. \revised{Recall that for the Pauli group of a single qubit, there are in total $ | \mathcal{P} |^2 =16 $ state pairs, which correspond to the sixteen nodes depicted in \refFigure{fig_pauli_1a}. More precisely, each of the nodes represents one of the $ 16 $ two-dimensional states $\{ (I,I), (I,X), (I,Y), (I,Z), (X,I),$ $ ..., (Z,Z)\}$. The initial state pair $(\rho_{0}, \sigma_{0})$, which here satisfies $ \rho_{0}=\sigma_{0}$, corresponds to state $1$ in \refFigure{fig_pauli_1a}.} The bad state pairs that constitute $\mathcal{B}_{\ket{\zeta_0},\delta}^{\ket{\zeta_0}}$ and $\mathcal{B}_{\ket{\xi_0},\delta}^{\ket{\xi_0}}$, which have a trace distance over $ \delta =1/5 $, are indicated in red. \lma{Each edge depicts the possibility of the two-dimensional Markov chain to jump between the two connected nodes.} Note that the number of bad state pairs can be affected by the choice of initial state. \refFigure{fig_pauli_1} shows the probability of maximum error in \eqref{eqn:Random_circuits__Maximum} and the maximum number of tolerant gates in \eqref{maximum_gates} for the same two different initial states: $\ket{\zeta_0}$ (upper) and $\ket{\xi_{0}}$ (bottom). \refFigure{fig_pauli_1a} and \refFigure{fig_pauli_1} illustrate too that the choice of initial state can affect the probability $ \mathbb{P}[\max_{0 \leq s \leq t} D_s > \delta] $ and the maximum number of tolerant gates $t_{\delta,\gamma}^\star$. Finally, when starting from the initial state $\ket{\zeta_0}$, in this simple case, \eqref{eqn:Random_circuits__Maximum} reduces to
\begin{equation*}
\mathbb{P}[\max\limits_{0 \leq s \leq t} D_s >1/5] 
= 1 - \bigl( 1 - \tfrac{2}{3} r \bigr)^t,
\end{equation*}
while when starting from the initial state $\ket{\xi_{0}}$ we have
\begin{equation*}
\mathbb{P}[\max\limits_{0 \leq s \leq t} D_s >1/5] =1-{(1-r)^{t}}.
\end{equation*}

\subsection{Error accumulation in nonrandom circuits}\label{section32}

\begin{figure}[hbt]
\begin{center}
\includegraphics[
width=0.85\linewidth
]%
{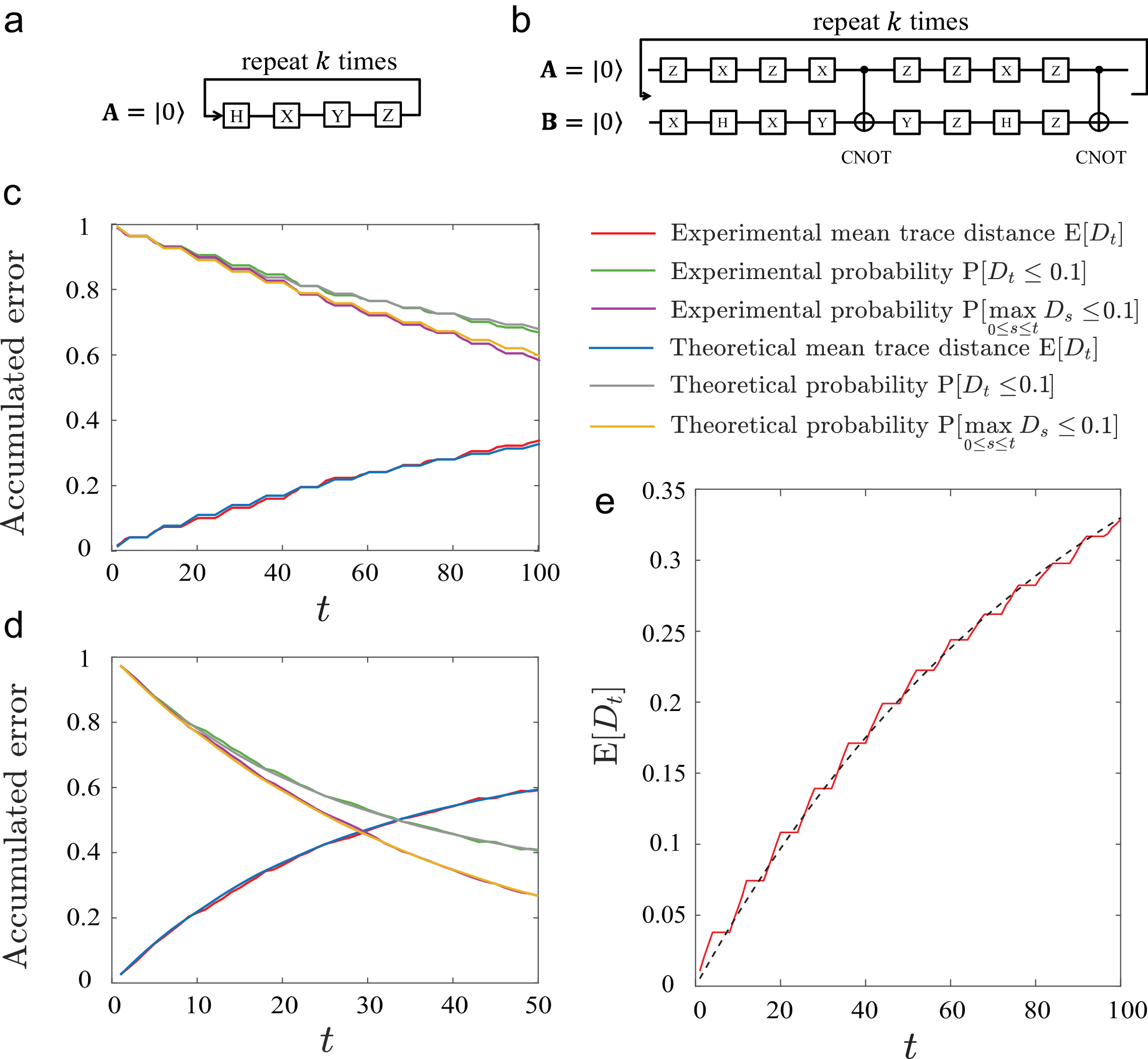}%
\caption{Theoretical and simulation results for error accumulation on a single-qubit circuit (figures a, c, and e) and a two-qubit circuit (figures b and d). The numerical results are calculated from 2000 independent runs, and almost indistinguishable from the formulae. \revised{The dashed, black curve in figure e is a fit of \eqref{equ:error_depolarizing_channel} to the data. The fit parameter is $ \mu^{\mathrm{fit}} \approx 0.011 $.}} 
\label{fig_fixed_circuit}
\end{center}
\end{figure}

Here we illustrate error accumulation rates in two nonrandom circuits. 
The first is a periodical single-qubit circuit that repeats a Hadamard, Pauli-$X$, Pauli-$Y$ and Pauli-$Z$ gate $ k=25 $ times, and the second a two-qubit circuit that is repeated $k = 5$ times; see also \refFigure{fig_fixed_circuit}. 
Here the controlled-NOT gate
$
   \mathrm{CNOT}
   =
   \bigl( 
      ( 1, 0, 0, 0 ); \allowbreak
      ( 0, 1, 0, 0 ); \allowbreak
      ( 0, 0, 0, 1 ); \allowbreak 
      ( 0, 0, 1, 0 ) 
   \bigr)
   .   
$
Consider also the following two error models in which the errors depend on the gates:

\noindent
(i) For the single-qubit circuit, presume
$
   \mathbb{P}[\Lambda=I ] = 0.990, \mathbb{P}[\Lambda=Z ] = 0.010.
$

\noindent
(ii) For the two-qubit circuit, when labeling the qubits by $ A $ and $ B $, suppose
\begin{gather}
   \mathbb{P}[\Lambda_A=I ] = 0.990, \, \mathbb{P}[\Lambda_A=X ] = 0.006,
   \mathbb{P}[\Lambda_A=Y ] = 0.003, \, \mathbb{P}[\Lambda_A=Z ] = 0.001;
   \nonumber \\ 
   \mathbb{P}[\Lambda_B=I ] = 0.980, \, \mathbb{P}[\Lambda_B=X ] = 0.002, 
   \mathbb{P}[\Lambda_B=Y ] = 0.014, \, \mathbb{P}[\Lambda_B=Z ] = 0.004. 
\end{gather}
In order to evaluate \refProposition{pro:nonrandom circuits}, we set the error threshold $\delta = 1/10$. 

The theoretical and simulation results on the two circuits are shown in \refFigure{fig_fixed_circuit}. Note that the simulation curves almost coincide with the theoretical curves; the deviation is only due to numerical limits. Furthermore, because different gates influence error accumulation to different degrees, the periodical ladder shape occurs in \refFigure{fig_fixed_circuit}. Observe furthermore that this periodical ladder shape is not captured by the fit method that only takes into account the decay of $t$ applications of a single depolarizing channel. 

~ 
\subsection{Continuous, random error accumulation in a single qubit}

\begin{figure}[!hbtp]
\begin{center}
\includegraphics[
width=0.7\linewidth
]%
{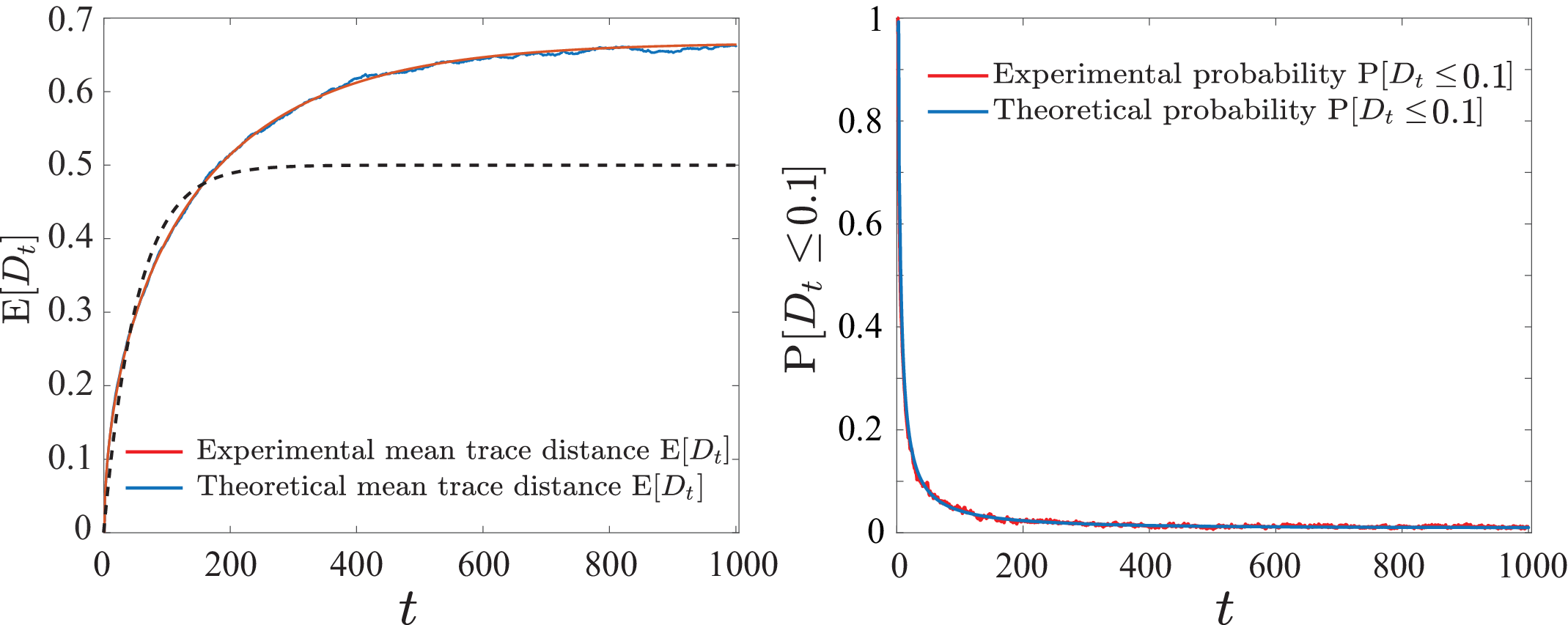}%
\caption{Continuous error accumulation in one qubit. The numerical results are from 2000 independent runs of our simulation. \revised{The dashed, black curve \lma{in the left figure} is a fit of \eqref{equ:error_depolarizing_channel} to the data. The resulting fit parameter is $ \mu^{\mathrm{fit}} \approx 0.019 $.}} 
\label{fig_contineous}%
\end{center}
\end{figure}

We now simulate the accumulation of continuous errors without depolarization ($ q=0 $) in a single qubit. Here, the noise is assumed to lead to a random walk on the Bloch sphere that takes steps of a fixed angle $ \alpha=1/10 $, and therefore $ p_t (\alpha)=\delta (\alpha) $. The threshold $ \delta $ is set to be $ 1/10 $. The theoretical mean trace distance  $ \expectation{D_t}$ and probability $ \probability{ D_t \leq \delta }$ are calculated using \eqref{eqn:Single_qubit__Expectation} and \eqref{eqn:Single_qubit__Distribution}. The theoretical results and simulations are shown in \refFigure{fig_contineous}. \revised{Note again that the traditional fit method disagrees at large $t$: this happens here because $\alpha \neq 0$.}

\section{Minimizing errors in quantum circuit through optimization}
\label{sec:Circuit_optimization}

The rate at which errors accumulate may be different for different quantum circuits that can implement the same algorithm. Using techniques from optimization and \eqref{eqn:Fixed_circuits__Maximum}, we can therefore search for the quantum circuit that has the lowest error rate accumulation while maintaining the same final state. To see this, suppose we are given a circuit
$
\mathcal{U}_\tau= \{ U_1, U_2, \ldots, U_\tau \}
$.
For given $\rho_0$ this brings the quantum state to some quantum state $\rho_\tau$. Other circuits may go to the same final state and have a lower probability of error at time $ \tau $. We will therefore aim to
\begin{equation}
    \label{minimize}
    \begin{aligned}
    & \underset{ G_1, \ldots, G_\tau \in \mathcal{G}_n }{\text{minimize}}
    & & u( \{ G_1, \ldots, G_\tau \} ) \\
    & \text{subject to}
    & & G_\tau \cdots G_1 = U_\tau \cdots U_1.
    \end{aligned}
\end{equation}
Here, one can for example choose for the objective function $u(\cdot)$ the probability of error \eqref{eqn:Fixed_circuits__Distribution}, or probability of maximum error \eqref{eqn:Fixed_circuits__Maximum}. To solve \eqref{minimize}, we design a simulated annealing algorithm in \refSection{sec:Simulated_annealing} to improve the quantum circuit.

The minimization problem in \eqref{minimize} is well-defined and has a few attractive features. For starters, the minimization problem automatically detects shorter circuits if the probability of error when applying the identity operator $I^{\otimes n}$ is relatively small. The optimum may then for example occur at a circuit of the form
\begin{equation}
        G_\tau G_{\tau-1} G_{\tau-2} \cdots G_2 G_1 
        = I^{ \otimes n } G_{\tau-1} I^{\otimes n} \cdots I^{\otimes n} G_1,
\end{equation}
which effectively means that only the two gates $G_{\tau-1} G_1$ are applied consecutively. The identity operators in this solution essentially describe the passing of time. Now, critically, note that while the minimization problem does consider all shorter circuits of depth at most $\tau$, this does not necessarily mean that the physical application of one specific group element $G \in \mathcal{G}_n$ is always the best. 
Concretely, in spite of the fact that any quantum circuit of the form $G_\tau \cdots G_1 = G \in \mathcal{G}_n$ performs the single group element $G \in \mathcal{G}_n$, it is not necessarily true that
\begin{equation}
    u( \{ G, I^{\otimes n}, \ldots, I^{\otimes n} \} )    
    <
    u( \{ G_1, \ldots, G_\tau \} )  
    .
\end{equation}
The reason for this is that the error distribution on the direct group element $G$ may be worse than using a circuit utilizing multiple other group elements. In other words, the optimal circuit need not always be the `direct' circuit, but of course it can be. (In \refSection{subsection:example} we also consider the situation in which an experimentalist can only apply a subset $\mathcal{A} \subseteq \mathcal{G}_n$ that need not necessarily be a group, and in such a case the direct group element $G$ may not even be a viable solution to the experimentalist if $G \not\in \mathcal{A}$.) Typically, the minimization problem will prefer shorter circuits if the probability of error when applying the identity operator $I^{\otimes n}$ is relatively small and the error distributions of all gate distributions are relatively homogeneous.

\begin{unabridged}
~ 
\subsection{Simulated annealing}
\label{sec:Simulated_annealing}

We will generate candidate circuits as follows. Let 
$
\{ \itr{G}{\eta}_1, \ldots, \itr{G}{\eta}_\tau \}
$
denote the circuit at iteration $\eta$. Choose an index $I \in [\tau-1]$ uniformly at random, choose $G \in \mathcal{G}$ uniformly at random. Then set
\begin{equation}
    \itr{G}{\eta+1}_i 
    =
    \begin{cases}
    G & \textrm{if } i = I, \\
    \itr{G}{\eta}_{I+1} \itr{G}{\eta}_{I} G^{\leftarrow} & \textrm{if } i = I + 1, \\
    \itr{G}{\eta}_i & \textrm{otherwise}. \\
    \end{cases}
\end{equation}
Here, $G^\leftarrow$ denotes the (left) inverse group element, i.e., $G^\leftarrow G = I^{\otimes n}$. The construction thus ensures that 
\begin{equation}
    \itr{G}{\eta+1}_{I+1} \itr{G}{\eta+1}_I 
    = \bigl( \itr{G}{\eta}_{I+1} \itr{G}{\eta}_{I} G^{\leftarrow} \bigr) G 
    = \itr{G}{\eta}_{I+1} \itr{G}{\eta}_{I}    
\end{equation}    
so that the circuit's intent does not change: $\itr{G}{\eta+1}_\tau \cdots \itr{G}{\eta+1}_1 = \itr{G}{\eta}_\tau \cdots \itr{G}{\eta}$.

We will use the Metropolis algorithm. 
Let
\begin{equation}
E 
= \bigl\{ \{ G_1, \ldots, G_\tau \} | G_\tau \cdots G_1 = U_\tau \cdots U_1 \bigr\}
\end{equation}
denote the set of all viable circuits. For two arbitrary circuits $i,j \in E$, let
\begin{equation}
\Delta(i,j) 
\triangleq \sum_{s=1}^{\tau-1} \indicator{ i_s \neq j_s, i_{s+1} \neq j_{s+1} }
\end{equation}
denote the number of consecutive gates that differ between both circuits. Under this construction, the \emph{candidate-generator matrix} of the Metropolis algorithm is given by
\begin{equation}
q_{ij} 
=
\begin{cases}
\frac{1}{ (\tau-1) \cardinality{ \mathcal{G} } } & \textrm{if } \Delta(i,j) \leq 1 \\
0 & \textrm{otherwise}. \\
\end{cases}
\end{equation}
Since the candidate-generator matrix is symmetric, this algorithm means that we set
$
\alpha_{i,j}(T)
= \exp{ \bigl( - \frac{1}{T} \max{ \{ 0, u(j) - u(i) \} } \bigr) }
$
as the \emph{acceptance probability} of circuit $j$ over $i$. Here $T \in(0,\infty)$ is a positive constant. Finally, we need a cooling schedule. Let
$
    M \triangleq \sup_{ \{ i,j \in E | \Delta(i,j) \leq 1 \} } \{ u(j) - u(i) \}.
$
Based on \cite{bremaud2017discrete}, if we choose a cooling schedule $\process{ T_\eta }{ \eta \geq 0 }$ that satisfies
$
    T_\eta 
    \geq 
    \frac{ \tau M }{ \ln{\eta} },
$
then the Metropolis algorithm will converge to the set of global minima of the minimization problem in \eqref{minimize}.

\begin{algorithm}[!hbtp]
    \footnotesize
    \KwIn{A group $\mathcal{G}$, a circuit $\{ U_1, \ldots, U_\tau \}$, and number of iterations $w$}
    \KwOut{A revised circuit $\{ \itr{G}{w}_1, \ldots, \itr{G}{w}_\tau \}$}
    \Begin{
    Initialize $\{ \itr{G}{0}_1, \ldots, \itr{G}{0}_\tau \} = \{ U_1, \ldots, U_\tau \}$\;
    \For{$\eta \leftarrow 1$ \KwTo $w$}{
    Choose $I \in [\tau-1]$ uniformly at random\;
    Choose $G \in \mathcal{G}$ uniformly at random\;
    Set $J_I = G, J_{I+1} = \itr{G}{\eta}_{I+1} \itr{G}{\eta}_{I} G^{\leftarrow}, J_i = \itr{G}{\eta}_i \, \forall_{i \neq I, I+1}$\;
    Choose $X \in [0,1]$ uniformly at random\;
    \uIf{$X \leq \alpha_{ \itr{G}{\eta},J }(T_\eta)$}{
    Set $\itr{G}{\eta+1} = J$\;
    }
    \Else{
    Set $\itr{G}{\eta+1} = \itr{G}{\eta}$\;
    }
    }
    }
    \caption{Pseudo-code for the simulated annealing algorithm described in \refSection{sec:Simulated_annealing}.}
    \label{pseudo:The_Simulated_Annealing_algorithm}
\end{algorithm}

\begin{lemma}
\label{lem:converge_to_global_minimize}
Algorithm \ref{pseudo:The_Simulated_Annealing_algorithm} converges to the global minimizer of \eqref{minimize} whenever $
T_\eta 
\geq { \tau M } / { \ln{\eta} }$ for $ \eta = 1, 2, \cdots $.
\end{lemma}

\subsection{Examples}\label{subsection:example}
\subsubsection{Gate-dependent error model}

We are going to improve the one-qubit circuit in \refFigure{fig_fixed_circuit} using Algorithm~\ref{pseudo:The_Simulated_Annealing_algorithm}. The gates are limited to the Clifford group $ {\mathcal{C}}_{1}$ and the errors will be limited to the Pauli channel. The error probabilities considered here are gate-dependent and can be found in \refAppendixSection{errormodel}. The cooling schedule used here will be set as $T_\eta = C/\ln{(\eta+1)}$, and the algorithm's result when using $ C=0.004 $ is shown in \refFigure{fig_circuit_optimization_one_qubit}. \refFigure{fig_circuit_optimization_one_qubit} illustrates that the improved circuit can indeed lower the error accumulation rate. The circuit with the lowest error accumulation rate that was found is shown in \refAppendixSection{optimized_circuit}.
\end{unabridged}

\begin{abridged}
\begin{SCfigure}
\includegraphics[
width=0.55\linewidth,
]%
{figure7_abridged.eps}%
\caption{Circuit optimization when using a tailor-made simulated annealing algorithm. Note that the probability of maximum error \eqref{eqn:Fixed_circuits__Maximum} decreases with the number of iterations $ \eta $. Here, we started the optimization from the two-qubit circuit in \refFigure{fig_fixed_circuit}.}%
\label{fig_circuit_optimization_greedy}%
\end{SCfigure}
\end{abridged}
\begin{unabridged}
\begin{figure}[!hbtp]
\begin{center}
\includegraphics[
width=0.5\linewidth
]%
{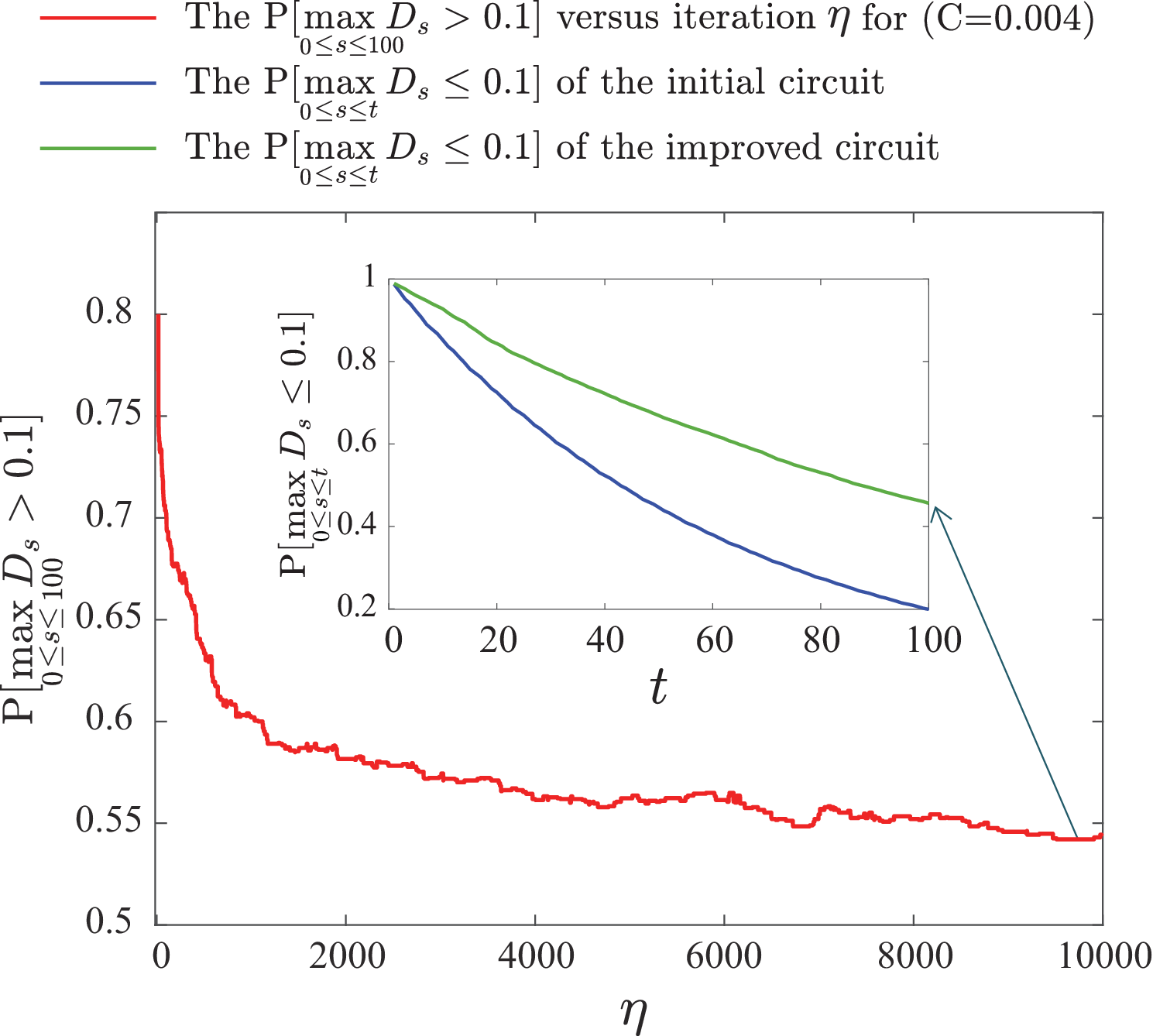}%
\caption{Circuit optimization when using Algorithm \ref{pseudo:The_Simulated_Annealing_algorithm}. The error probabilities are gate-dependent. Note that the probability of maximum error \eqref{eqn:Fixed_circuits__Maximum} decreases as the number of iterations $ \eta $ increases when using Algorithm \ref{pseudo:The_Simulated_Annealing_algorithm} ($C=0.004$). Here we started from the one-qubit circuit in \refFigure{fig_fixed_circuit}.}%
\label{fig_circuit_optimization_one_qubit}%
\end{center}
\end{figure}
\end{unabridged}

\begin{unabridged}

\subsubsection{Gates in a subset of one group}
\label{sec:Gates_in_a_subset_of_one_group}

The gates that are available in practice may be restricted to some subset $ \mathcal{A}\subseteq \mathcal{G} $ not necessarily a group. Under such constraint, we could generate candidate circuits as follows: Let $ \{ \itr{G}{\eta}_1, \ldots, \itr{G}{\eta}_\tau \} $ denote the circuit at iteration $\eta$. In each iteration, two neighboring gates will be considered to be replaced by two other neighboring gates. There are $ m\leq (\tau-1) $ neighboring gate pairs $ (\itr{G}{\eta}_1, \itr{G}{\eta}_{2}), \ldots, (\itr{G}{\eta}_{m-1}, \itr{G}{\eta}_{m})$ that can be replaced by two different neighboring gates. Choose an index $I \in [m-1]$ uniformly at random, and replace $(\itr{G}{\eta}_I, \itr{G}{\eta}_{I+1})$ by any gate pair from $ \{(\tilde{G_1}, \tilde{G_2})\in \mathcal{A}^2 \mid \itr{G}{\eta}_I\itr{G}{\eta}_{I+1}=\tilde{G_1}\tilde{G_2}\}$ uniformly at random. Pseudo-code for this modified algorithm can be found in \refAppendixSection{appendix:Pseudo_code_for_gate_limited_simulated_annealing}. It must be noted that this algorithm is not guaranteed to converge to the global minimizer of \eqref{minimize} (due to limiting the gates available); however, it may still find use in practical scenarios where one only has access to a restricted set of gates.

We now aim to decrease the probability of maximum error \eqref{eqn:Fixed_circuits__Maximum} by changing the two-qubit circuit shown in Figure \ref{fig_fixed_circuit}. The error model is the same as that in \refSection{sec:Numerical_results}--B. The set of gates available for improving the circuit is here limited to $ \lbrace I, X, Y, Z, H, CNOT\rbrace $. The result here for the two-qubit circuit is obtained by again using the cooling schedule $T_\eta = C/\ln{(\eta+1)}$ but now letting the parameter $ C=0.002 $. Figure \ref{fig_circuit_optimization_greedy} shows that a more error-tolerant circuit can indeed be found using this simulated annealing algorithm. The improved circuit is shown in \refAppendixSection{optimized_circuit}.

\begin{figure}[!hbtp]
\begin{center}
\includegraphics[
width=0.5\linewidth
]%
{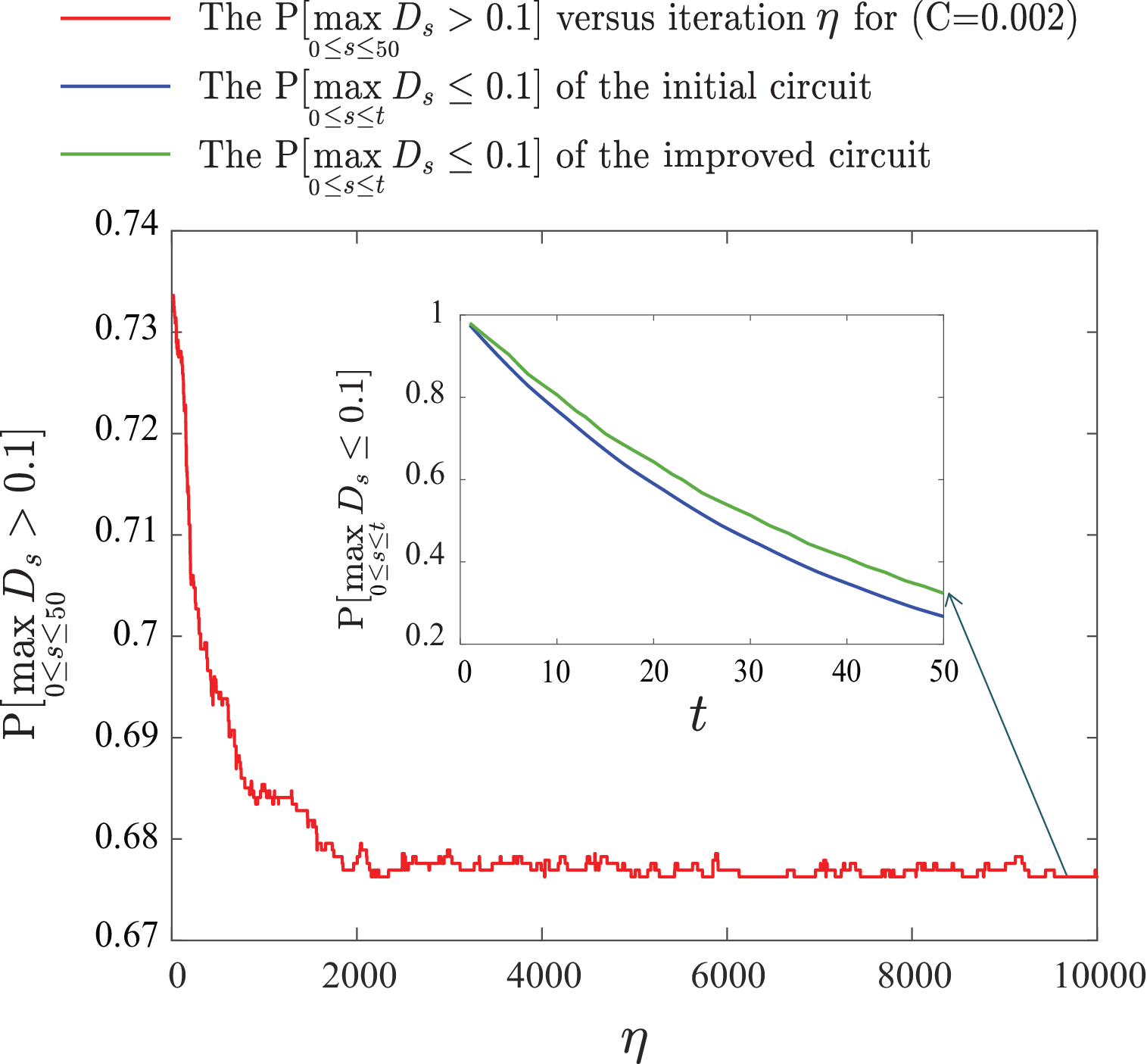}%
\caption{Circuit optimization when using Algorithm \ref{pseudo:Gate_limited_simulated_annealing}. The set of gates available is chosen limited to $ \lbrace I, X, Y, Z, H, CNOT\rbrace $. Note that the probability of maximum error \eqref{eqn:Fixed_circuits__Maximum} decreases as the number of iterations $ \eta $ increases when using Algorithm~\ref{pseudo:Gate_limited_simulated_annealing} ($C=0.002$). Here we started from the two-qubit circuit shown in Figure~\ref{fig_fixed_circuit}.}%
\label{fig_circuit_optimization_greedy}%
\end{center}
\end{figure}

\subsubsection{Deutsch--Jozsa Algorithm}

Let us give further proof of concept through the Deutsch--Jozsa Algorithm for one classical bit \cite{deutsch1992rapid,cleve1998quantum}. This quantum algorithm determines if a function $f : \{ 0, 1 \} \to \{ 0, 1 \}$ is constant or balanced, i.e., if $f(0) = f(1)$ or $f(0) \neq f(1)$. It is typically implemented using the quantum circuit in \refFigure{fig:DeutschJozsa_Algorithm_as_a_quantum_circuit}. If no errors occur in this quantum circuit, then the first qubit would measure $\ket{0}$ or $\ket{1}$ w.p.\ one if $f$ constant or balanced, respectively. If errors occur in this quantum circuit, then there is a strictly positive probability that the first qubit measures $\ket{1}$ or $\ket{0}$ in spite of $f$ being constant or balanced, respectively, and thus for the algorithm to incorrectly output that $f$ is constant or balanced. This \emph{misclassification probability} $\nu$ of the algorithm depends on the underlying error distributions, and can be calculated by adapting \eqref{eqn:Fixed_circuits__Distribution}'s derivation.

\begin{figure}[!hbtp]
\begin{center}
\includegraphics[
width=0.5\linewidth
]%
{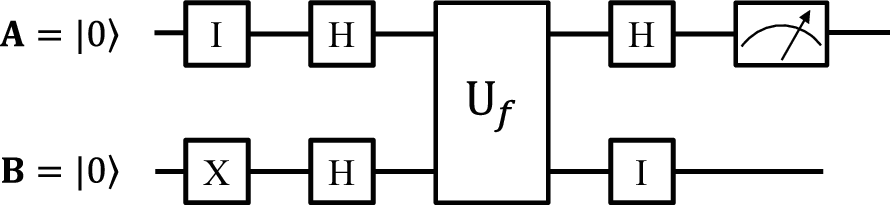}%
\caption{\revised{The Deutsch--Jozsa Algorithm for one classical bit in quantum circuit form.}}%
\label{fig:DeutschJozsa_Algorithm_as_a_quantum_circuit}
\end{center}
\end{figure}

We suppose now that errors occur according to a distribution in which two-qubit Clifford gates are more error prone than single-qubit gates, see \refAppendixSection{appendix:Error_distribution} for the details. We can then revise the quantum circuit in \refFigure{fig:DeutschJozsa_Algorithm_as_a_quantum_circuit} using a simulated annealing algorithm in \refAppendixSection{appendix:simulated_annealing_3} that aims at minimizing \eqref{minimize} by randomly swapping out poor gate pairs for better gate pairs. This simulated annealing algorithm, like any other, is sensitive to the choice of \emph{cooling schedule} \cite{bremaud2017discrete}, here set as
$
T_\eta = C \bigl( \gamma / \eta + (1-\gamma) / \ln{(\eta+1)} \bigr)
$
with $C > 0$, $\gamma \in [0,1]$; the integer $\eta$ indexes the iterations. \refFigure{fig:Heatmap_for_DeutschJozsas_Algorithm} shows the ratio $\Theta \triangleq \nu_{\textrm{original circuit}} / \nu_{\textrm{revised circuit}}$ as a function of $C, \gamma$ for $f_a(x)=x, f_b(x)=1-x, f_c(x)=0, f_d(x)=1$ where $x\in\{0,1\}$. Note that $\Theta \geq 1$ always, $\geq 1.60$ commonly, and sometimes even $\geq 2.20$. 

\begin{figure}[!hbtp]
\begin{center}
\includegraphics[
width=0.5\linewidth
]%
{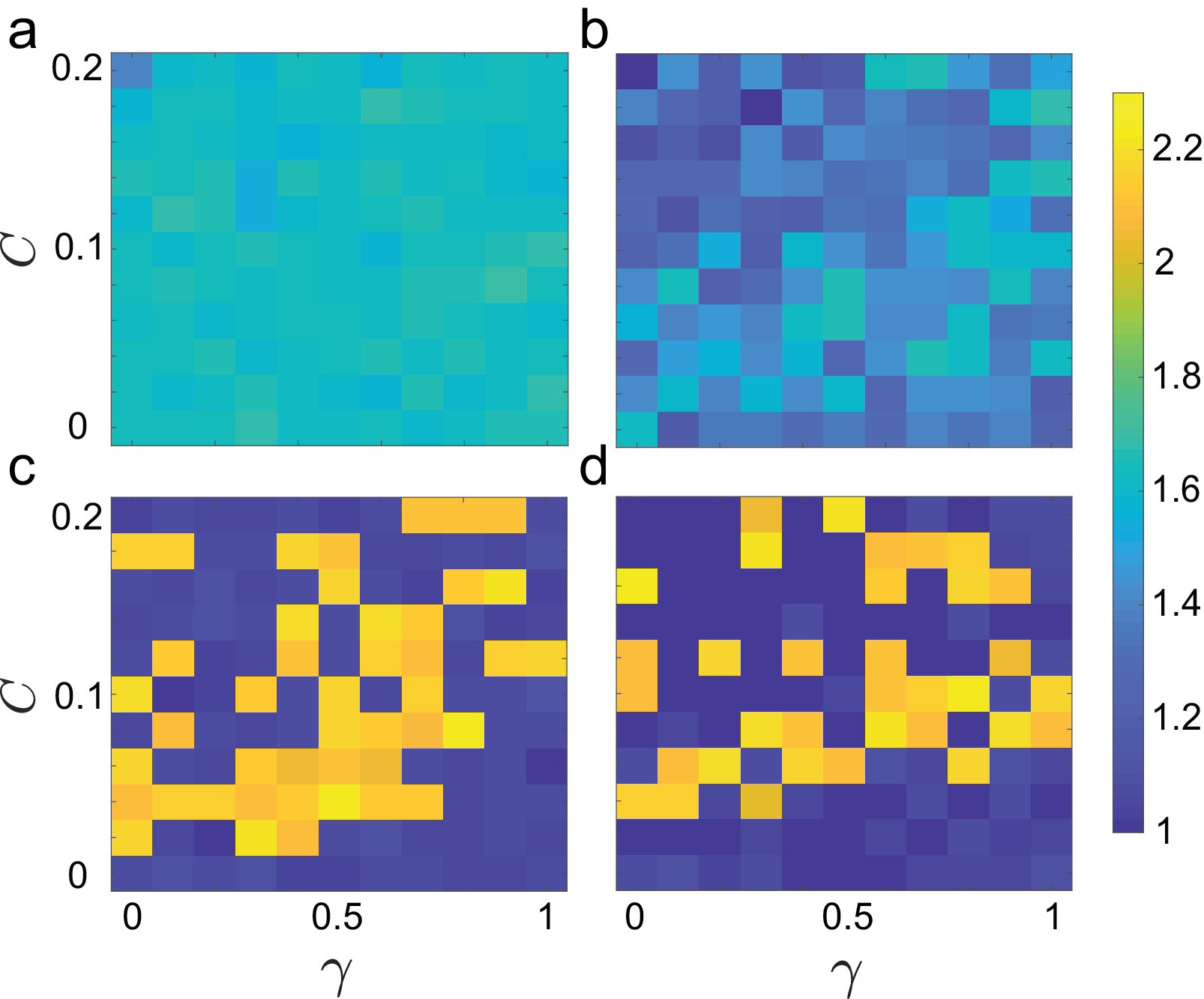}%
\caption{
\revised{For every pair $(C,\gamma)$ here, $\Theta$ was calculated using a Monte Carlo simulation with $ 10^5 $ independent repetitions for the best circuit found throughout $w = 10^3$ iterations of the annealing algorithm. $u(\cdot)$ was set to the misclassification probability for a, c; and to \eqref{eqn:Fixed_circuits__Maximum} for b, d.}
}%
\label{fig:Heatmap_for_DeutschJozsas_Algorithm}%
\end{center}
\end{figure}

\end{unabridged}
\section{Conclusion}
\label{sec:Conclusion}

In conclusion; we have proposed and studied a model for discrete Markovian error accumulation in a multi-qubit quantum computation, as well as a model describing continuous errors accumulating in a single qubit. By modeling the quantum computation with and without errors as two coupled Markov chains, we were able to capture a weak form of time-dependency, allow for fairly generic error distributions, and describe multi-qubit systems. Furthermore, by using techniques from discrete probability theory, we could calculate the probability that error measures such as the fidelity and trace distance exceed a threshold analytically. To combat the numerical challenge that may occur when evaluating our expressions, we additionally provided an analytical bound on the error probabilities that is of lower numerical complexity. Finally, we showed how our expressions can be used to decide how many gates one can apply before too many errors accumulate with high probability, and how one can lower the rate of error accumulation in existing circuits by using techniques from optimization. 

This paper lay down a foundation for one error accumulation model, and multiple interesting follow-up topics can now be investigated as future research.
Here, we provide \revised{five} intriguing ideas:

\noindent
-- The accumulation of errors when using a universal gate set would need to be modeled using stochastic processes that live on infinite state spaces. Such an approach looks to be connected to the modeling of random walks on manifolds. This would be a challenging, intriguing, and important next step for the analysis of error accumulation in quantum circuits.

\noindent
-- The expressions in \eqref{eqn:Random_circuits__Maximum} and \eqref{eqn:Fixed_circuits__Maximum} are, essentially, generalized forms of a geometric distribution. For particular groups and error models, it may be that this expression is well-approximated by a standard geometric distribution (which would be of substantially lower numerical complexity). It would be interesting to investigate whether a reduction of  \eqref{eqn:Random_circuits__Maximum} and \eqref{eqn:Fixed_circuits__Maximum} occurs, or whether an approximation can be found, for particular quantum systems. 

\noindent
-- With that idea in mind, note that the hitting time of the set $\mathcal{B}_{\ket{\psi_0},\delta}^{\ket{\Psi_0}}$ is naturally related to its size relative to the size of the group $\mathcal{G}_n$. As the number of qubits increases, both of these sets grow in size. Investigating the growth relation between these two sets for particular groups via e.g.\ techniques from analytical combinatorics \cite{flajolet2015analytic} may reveal an asymptotic distributional law for the errors in quantum computations with many qubits.

\noindent
-- The availability of an analytical expression for the accumulation of errors allows us to proceed with second-tier optimization methods. For example, any quantum computer architecture would, to achieve practical quantum computing in the near future, have some classical control mechanism that routinely takes operational decisions: which gate do we apply next, do we now apply an error correction procedure, etc. Each of these different operations has its own cost associated with it, e.g.\ in the form of classical compute time or the loss of ancillary qubits. Using techniques from decision theory \cite{puterman1994markov}, we can weigh the long-term effects of different operations through the available analytical expressions, and we could overall achieve more efficient computations in the future. Essentially, we could then compute more with fewer qubits.

\revisedPartBegin
\noindent
-- It would be interesting future research to appropriately scale space and time of our coupled two-dimensional Markov chain, and examine the stochastic differential equation that would emerge. Because our model is a hybrid of classical probability theory and quantum information theory, we expect that the stochastic differential equation that will emergence will not be a stochastic Schr\"{o}dinger equation. Instead we expect a diffusion process on a manifold, which is something that \refProposition{prop:Error_accumulation_in_a_single_qubit} also hints at. What could valuably come from this direction of research would be a numerically more efficient tool for error calculations in larger quantum circuits.
\revisedPartEnd

\paragraph*{Acknowledgments.}
We are grateful to Bart van Schooten, who contributed the code on TU/e's GitLab server. Finally, this research received financial support from the Chinese Scholarship Council (CSC) in the form of a CSC Scholarship.

\bibliography{bibitem}
\bibliographystyle{splncs04}

\appendix
\section{$\mathcal{C}_n$ is a group}
\label{appendix:group}

The fact that $\mathcal{C}_n$ is a group can be verified by checking the necessary properties:

\revisedPartBegin
\noindent
\emph{Binary operation.} Suppose $A, B \in \mathcal{C}_n$. Thus for all $\sigma \in \pm P_n^*$, $A \sigma A^\dagger \in \pm P_n^*$ and $B \sigma B^\dagger \in \pm P_n^*$; and moreover $A (B \sigma B^\dagger) A^\dagger \in \pm P_n^*$. Let $\sigma \in \pm P_n^*$ be arb., and note that we have shown that $(AB) \sigma (AB)^\dagger \in \pm P_n^*$. Thus $A B \in \mathcal{C}_n$. 

\noindent
\emph{Associativity.} This is for free because matrix multiplication is associative.
\revisedPartEnd

\noindent
\emph{Identity.} $I^{ \otimes n } \in \mathcal{C}_n$ because it is unitary and for all $\sigma \in \pm P_n^*$, $I^{ \otimes n } \sigma ( I^{ \otimes n } )^\dagger = \sigma$. 

\noindent
\emph{Inverses.} Suppose $C \in \mathcal{C}_n$, such that for any $\sigma \in \pm P_n^*$ we have that $C \sigma C^\dagger \in \pm P_n^*$. This implies that for any $\omega \in \pm P_n^*$, we can find a $\sigma \in \pm P_n^*$ such that $\omega = C \sigma C^\dagger$ (isomorpishm). Conclude that because $C$ is unitary, $C^{-1} \omega ( C^{-1} )^\dagger = C^\dagger \omega C = C^\dagger C \sigma C^\dagger C = \sigma \in P_n^*$. Hence $C^{-1} \in \mathcal{C}_n$. 
\qed

\section{Relation between the error probabilities when using the trace distance and fidelity}
\label{appendix:Relation_between_the_error_probabilities_when_using_the_trace_distance_and_fidelity}

Let $ t $ be such that $0\leq t \leq \tau$ and let $\omega \in \{ D_t \leq \varepsilon \} = \{ 1 - D_t \geq 1 - \varepsilon \}$. By \cite[(9.110)]{nielsen_quantum_2011}, we have that $1 - F_t \leq D_t \leq \sqrt{ 1 - F_t^2 }$ for all $t \geq 0$. Consequentially $1 - D_t \leq F_t \leq \sqrt{1 - D_t^2 }$ for all $t \geq 0$. On every such $\omega$, we thus also have that $F_t \geq 1 - \varepsilon$. We have shown that $\{ D_t \leq \varepsilon \} \subseteq \{ F_t \geq 1 - \varepsilon \}$, which proves the first statement. For the second statement, we similarly note that
$
\{ \min_{0 \leq s \leq t} F_s \geq 1 - \varepsilon \}
\supseteq \{ \min_{0 \leq s \leq t} ( 1 - D_s ) \geq 1 - \varepsilon \}
=\{ \max_{0 \leq s \leq t} D_s \leq \varepsilon \}
$. \qed

\revisedPartBegin

\section{An example explicit calculation of the results in \refProposition{prop:Error_accumulation_in_a_single_qubit} and \refLemma{lemma2}}
\label{appendix:example_to_calculate_the_lower_bound}

In order to calculate the results of \refProposition{prop:Error_accumulation_in_a_single_qubit} or \refLemma{lemma2}, one requires a transition matrix $P$. For the example of randomized benchmarking in \refSection{sec:Error_accumulation_in_randomized_benchmarking} with Pauli channels $\{ I, X, Y, Z \}$ only, when enumerating the two-dimensional states
\begin{equation}
\mathcal{G}_1^2 
= \bigl\{ (I,I), (I,X), (I,Y), (I,Z), (X,I), (X,X), ..., (Z,Z) \bigr\}
\end{equation}
lexicographically along both the rows and columns (so as indicated), the transition matrix $ P $ is represented as follows:
\begin{equation}
P =
\begin{bsmallmatrix}
\tfrac{1-r}{4} & \tfrac{r}{12} &\tfrac{r}{12} & \tfrac{r}{12}  &  \tfrac{r}{12} & \tfrac{1-r}{4}  & \tfrac{r}{12} & \tfrac{r}{12}  & \tfrac{r}{12}  &  \tfrac{r}{12}   & \tfrac{1-r}{4}  & \tfrac{r}{12}  & \tfrac{r}{12} &\tfrac{r}{12} & \tfrac{r}{12} & \tfrac{1-r}{4}\\
 \tfrac{r}{12}  & \tfrac{1-r}{4} &  \tfrac{r}{12}  &  \tfrac{r}{12}  & \tfrac{1-r}{4} &\tfrac{r}{12} & \tfrac{r}{12}  &  \tfrac{r}{12}  &\tfrac{r}{12} & \tfrac{r}{12}  &  \tfrac{r}{12} & \tfrac{1-r}{4}  & \tfrac{r}{12} & \tfrac{r}{12}  & \tfrac{1-r}{4} & \tfrac{r}{12}\\
\tfrac{r}{12}  & \tfrac{r}{12} & \tfrac{1-r}{4}  & \tfrac{r}{12}  & \tfrac{r}{12} & \tfrac{r}{12}  & \tfrac{r}{12} & \tfrac{1-r}{4}  & \tfrac{1-r}{4}  & \tfrac{r}{12}  & \tfrac{r}{12} & \tfrac{r}{12}  & \tfrac{r}{12} & \tfrac{1-r}{4}  & \tfrac{r}{12} & \tfrac{r}{12}\\
\tfrac{r}{12}  & \tfrac{r}{12} & \tfrac{r}{12} & \tfrac{1-r}{4}  &  \tfrac{r}{12} &  \tfrac{r}{12}  & \tfrac{1-r}{4} &  \tfrac{r}{12}  &  \tfrac{r}{12}  & \tfrac{1-r}{4}  &  \tfrac{r}{12} &  \tfrac{r}{12}  & \tfrac{1-r}{4} &  \tfrac{r}{12}  &  \tfrac{r}{12} &  \tfrac{r}{12}\\
\tfrac{r}{12}  & \tfrac{1-r}{4} & \tfrac{r}{12}  & \tfrac{r}{12}  & \tfrac{1-r}{4} & \tfrac{r}{12}  & \tfrac{r}{12} & \tfrac{r}{12}  & \tfrac{r}{12}  & \tfrac{r}{12}  & \tfrac{r}{12} & \tfrac{1-r}{4}  & \tfrac{r}{12} & \tfrac{r}{12}  & \tfrac{1-r}{4} & \tfrac{r}{12}\\
\tfrac{1-r}{4}  & \tfrac{r}{12} & \tfrac{r}{12}  & \tfrac{r}{12}  & \tfrac{r}{12} & \tfrac{1-r}{4}  & \tfrac{r}{12} & \tfrac{r}{12}  & \tfrac{r}{12}  & \tfrac{r}{12}  &  \tfrac{1-r}{4} & \tfrac{r}{12}  & \tfrac{r}{12} & \tfrac{r}{12}  & \tfrac{r}{12} &  \tfrac{1-r}{4}\\
\tfrac{r}{12}  & \tfrac{r}{12} & \tfrac{r}{12}  &  \tfrac{1-r}{4}  & \tfrac{r}{12} & \tfrac{r}{12}  & \tfrac{1-r}{4} & \tfrac{r}{12}  & \tfrac{r}{12}  &  \tfrac{1-r}{4}  & \tfrac{r}{12} & \tfrac{r}{12}  &  \tfrac{1-r}{4} & \tfrac{r}{12}  & \tfrac{r}{12} & \tfrac{r}{12}\\
\tfrac{r}{12}  & \tfrac{r}{12} &  \tfrac{1-r}{4}  & \tfrac{r}{12}  & \tfrac{r}{12} & \tfrac{r}{12}  & \tfrac{r}{12} & \tfrac{1-r}{4}  &  \tfrac{1-r}{4}  & \tfrac{r}{12}  & \tfrac{r}{12} & \tfrac{r}{12}  & \tfrac{r}{12} &  \tfrac{1-r}{4}  & \tfrac{r}{12} & \tfrac{r}{12}\\
\tfrac{r}{12}  & \tfrac{r}{12} &  \tfrac{1-r}{4}  & \tfrac{r}{12}  & \tfrac{r}{12} & \tfrac{r}{12}  & \tfrac{r}{12} & \tfrac{1-r}{4}  &  \tfrac{1-r}{4}  & \tfrac{r}{12}  & \tfrac{r}{12} & \tfrac{r}{12}  & \tfrac{r}{12} &  \tfrac{1-r}{4}  & \tfrac{r}{12} & \tfrac{r}{12}\\
\tfrac{r}{12}  & \tfrac{r}{12} & \tfrac{r}{12} & \tfrac{1-r}{4}  &  \tfrac{r}{12} &  \tfrac{r}{12}  & \tfrac{1-r}{4} &  \tfrac{r}{12}  &  \tfrac{r}{12}  & \tfrac{1-r}{4}  &  \tfrac{r}{12} &  \tfrac{r}{12}  & \tfrac{1-r}{4} &  \tfrac{r}{12}  &  \tfrac{r}{12} &  \tfrac{r}{12}\\
\tfrac{1-r}{4}  & \tfrac{r}{12} & \tfrac{r}{12}  & \tfrac{r}{12}  & \tfrac{r}{12} & \tfrac{1-r}{4}  & \tfrac{r}{12} & \tfrac{r}{12}  & \tfrac{r}{12}  & \tfrac{r}{12}  &  \tfrac{1-r}{4} & \tfrac{r}{12}  & \tfrac{r}{12} & \tfrac{r}{12}  & \tfrac{r}{12} &  \tfrac{1-r}{4}\\
\tfrac{r}{12}  & \tfrac{1-r}{4} & \tfrac{r}{12}  & \tfrac{r}{12}  & \tfrac{1-r}{4} & \tfrac{r}{12}  & \tfrac{r}{12} & \tfrac{r}{12}  & \tfrac{r}{12}  & \tfrac{r}{12}  & \tfrac{r}{12} & \tfrac{1-r}{4}  & \tfrac{r}{12} & \tfrac{r}{12}  & \tfrac{1-r}{4} & \tfrac{r}{12}\\
\tfrac{r}{12}  & \tfrac{r}{12} & \tfrac{r}{12}  &  \tfrac{1-r}{4}  & \tfrac{r}{12} & \tfrac{r}{12}  & \tfrac{1-r}{4} & \tfrac{r}{12}  & \tfrac{r}{12}  &  \tfrac{1-r}{4}  & \tfrac{r}{12} & \tfrac{r}{12}  &  \tfrac{1-r}{4} & \tfrac{r}{12}  & \tfrac{r}{12} & \tfrac{r}{12}\\
\tfrac{r}{12}  & \tfrac{r}{12} &  \tfrac{1-r}{4}  & \tfrac{r}{12}  & \tfrac{r}{12} & \tfrac{r}{12}  & \tfrac{r}{12} & \tfrac{1-r}{4}  &  \tfrac{1-r}{4}  & \tfrac{r}{12}  & \tfrac{r}{12} & \tfrac{r}{12}  & \tfrac{r}{12} &  \tfrac{1-r}{4}  & \tfrac{r}{12} & \tfrac{r}{12}\\
\tfrac{r}{12}  & \tfrac{1-r}{4} & \tfrac{r}{12}  & \tfrac{r}{12}  & \tfrac{1-r}{4} & \tfrac{r}{12}  & \tfrac{r}{12} & \tfrac{r}{12}  & \tfrac{r}{12}  & \tfrac{r}{12}  & \tfrac{r}{12} & \tfrac{1-r}{4}  & \tfrac{r}{12} & \tfrac{r}{12}  & \tfrac{1-r}{4} & \tfrac{r}{12}\\
\tfrac{1-r}{4} & \tfrac{r}{12} &\tfrac{r}{12} & \tfrac{r}{12}  &  \tfrac{r}{12} & \tfrac{1-r}{4}  & \tfrac{r}{12} & \tfrac{r}{12}  & \tfrac{r}{12}  &  \tfrac{r}{12}   & \tfrac{1-r}{4}  & \tfrac{r}{12}  & \tfrac{r}{12} &\tfrac{r}{12} & \tfrac{r}{12} & \tfrac{1-r}{4}
\end{bsmallmatrix}
\label{eqn:Example_P_matrix_for_randomized_benchmarking_with_Pauli_channels}
\end{equation}

\paragraph{Probability distribution of the maximum trace distance.}
Using \eqref{eqn:Example_P_matrix_for_randomized_benchmarking_with_Pauli_channels}, we can evaluate \refProposition{prop:Error_accumulation_distributions_for_random_circuits}'s results. When starting from the initial state $\ket{\zeta_0}$, \eqref{eqn:Random_circuits__Maximum} simplifies (after some algebra) to
\begin{equation}
\mathbb{P}[\max\limits_{0 \leq s \leq t} D_s >1/5] 
= 1 - \bigl( 1 - \tfrac{2}{3} r \bigr)^t.
\end{equation}
Similarly when starting from the initial state $\ket{\xi_{0}}$, \eqref{eqn:Random_circuits__Maximum} leads to
\begin{equation}
\mathbb{P}[\max\limits_{0 \leq s \leq t} D_s >1/5] =1-{(1-r)^{t}}.
\end{equation}

\paragraph{Lower bound.}
Using \eqref{eqn:Example_P_matrix_for_randomized_benchmarking_with_Pauli_channels}, we can also evaluate the lower bound in \refLemma{lemma2}. When the initial state $\ket{\zeta_0}= \sqrt{7/10}\ket{0}+ \sqrt{3/10}\ket{1}$, the expected hitting time of $\mathcal{B}_{\ket{\psi_0},1/5}^{\ket{\Psi_0}}$ turns out to be given by $\expectationWrt{ T_{ \mathcal{B}_{\ket{\psi_0},1/5}^{\ket{\Psi_0}} } }{z} = 3/(2r)$. The lower bound in \eqref{eqn:Lower_bound_on_the_maximum_probability_for_random_circuits} therefore reads 
\begin{equation}
\probabilityWrt{ \max_{ 0 \leq s \leq t } D_s > 1/5}{z_0}
\geq 0 \vee \Bigl( 1 - \frac{3}{2r(t + 1) } \Bigr).
\label{eqn:Lower_bound_Pauli_1}
\end{equation}
Here $ a\vee b\triangleq \max\{a, b\} $. Alternatively, when the initial state $\ket{\xi_{0}}= \sqrt{4/5}\ket{0}+ \sqrt{1/5}\ket{1}$, the expected hitting time of $\mathcal{B}_{\ket{\psi_0},1/5}^{\ket{\Psi_0}}$ can be calculated to be $\expectationWrt{ T_{ \mathcal{B}_{\ket{\psi_0},1/5}^{\ket{\Psi_0}} } }{z} = 1/r$. The lower bound is thus given by
\begin{equation}
\probabilityWrt{ \max_{ 0 \leq s \leq t } D_s > 1/5}{z_0}
\geq 0 \vee \Bigl( 1 - \frac{1}{r(t + 1) } \Bigr).
\label{eqn:Lower_bound_Pauli_1}
\end{equation}

\begin{figure}[!hbtp]
\begin{center}
\includegraphics[
width=0.8\linewidth
]%
{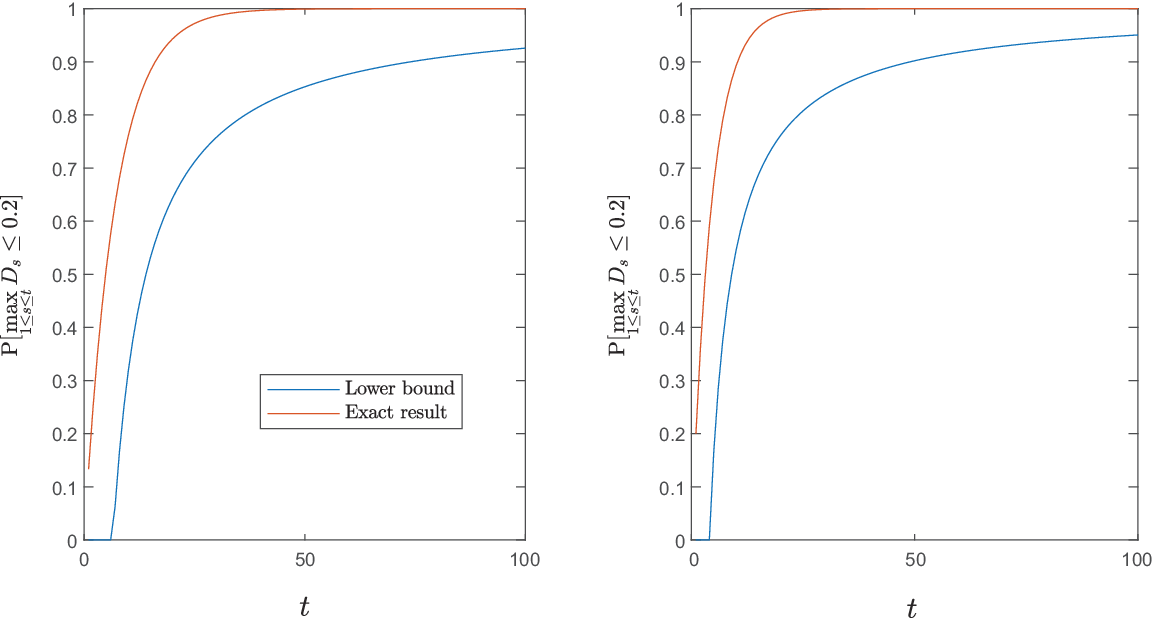}%
\caption{\revised{Lower bounds and exact probabilities $ \mathbb{P}[\max\limits_{0 \leq s \leq t} D_s >1/5] $ with $ r=0.2 $ for initial state $\ket{\zeta_0}$ (left) and $\ket{\xi_{0}}$ (right).}}%
\label{fig_lower_bound}%
\end{center}
\end{figure}

\paragraph{Comparison of the probability distribution of the maximum trace distance to its lower bound} The lower bounds and exact results with $ r=0.2 $ are shown in Figure \ref{fig_lower_bound}.

\paragraph{On how to construct a $P$ or $Q$ matrix.}
To assist you in constructing a transition matrix $P$ or $Q$, which are needed for the results in \refSection{sec:Model}, we have written an \texttt{R} script that can generate such matrices. The script generates a transition matrix when you a scenario with Pauli and Clifford channels and with error probabilities that are either dependent or independent of the gate: all you as user have to do, is to input a vector of (gate-dependent) error probabilities. The code of this script can be found at \url{https://gitlab.tue.nl/20061069/markov-chains-for-error-accumulation-in-quantum-circuits}. Additionally, for as long as the following public service remains available, the script can be tried out at \url{https://bevanschooten.shinyapps.io/qbiterrors/}.
\revisedPartEnd

\section{Number of stabilizer states for a gate}
\label{appendix:Number_of_stabilizer_states}

For $ n $ qubits, any gate $\mathcal{M}\in \mathcal{G}_{n}\setminus I^{\otimes n} $ can be represented using a $ 2^{n}\times 2^{n} $ unitary matrix. Recall that any unitary matrix of finite size is unitarily diagonalizable since every unitary matrix is normal~\cite{roman2005advanced}. A $ 2^{n}\times 2^{n} $ matrix that is diagonalizable must have a set of $2^{n}$ linearly independent eigenvectors~\cite{roman2005advanced}.

The initial states $ \ket{\psi_{0}} $ that can satisfy $ \mathcal{M}{|\psi_{0}}\rangle=e^{i\gamma}{|\psi_{0}}\rangle$ are the eigenvectors of the matrix $\mathcal{M} $ with eigenvalue $ \lambda=e^{i\gamma} $. For any unitary matrix $ A $ with eigenvalue $ \lambda $ and eigenvector $ v $, $ A^{\dagger}A=AA^{\dagger}=I $, $  v^{\dagger}v=v^{\dagger}A^{\dagger}Av=\lambda^{\dagger}v^{\dagger}v\lambda= \lambda^{\dagger}\lambda v^{\dagger}v$. Also recall that any eigenvector $ ||v||\neq 0 $ by definition~\cite{roman2005advanced} and thus it always holds that $ |\lambda|=1 $. So $ \mathcal{M}\ket{\psi_{0}}=\lambda\ket{\psi_{0}}=e^{i\gamma}\ket{\psi_{0}} $. \qed

\section{A stabilizer state follows after a stabilizer state}
\label{appendix:A_stabilizer_state_follows_after_a_stabilizer_state}

By assumption and the definition in \refEquation{reachable_state}, for any state $ \ket{\psi_{1}}\in{\mathcal{R}_{\ket{\psi_{0}}}} $, $ \exists \mathcal{Z}\in \mathcal{G}_n : \ket{\psi_{1}}=\mathcal{Z}\ket{\psi_{0}}$ since $ \mathcal{G}_n $ is a group. we have furthermore that $ \exists \mathcal{H}\in \mathcal{G}_n \setminus I^{\otimes n}: \mathcal{HZ}=\mathcal{ZM}$. Then $ \ket{\psi_{1}}=\mathcal{Z}\ket{\psi_{0}}=e^{-i\gamma}\mathcal{ZM}\ket{\psi_{0}}=e^{-i\gamma}\mathcal{HZ}\ket{\psi_{0}}=e^{-i\gamma}\mathcal{H}\ket{\psi_{1}}$. So $ \mathcal{H}\ket{\psi_{1}}=e^{i\gamma}\ket{\psi_{1}}$. \qed

\section{Gate-dependent error model}
\label{errormodel}

In \refTable{table:Specific_error_probabilities}, we provide the precise error probabilities used in \refSection{sec:Numerical_results}. The specific values were simply randomly generated to result in an inhomogeneous example; we spent no time post-selecting these values.

\begin{table}[!hbtp]
\caption{The specific error probabilities used in \refSection{sec:Circuit_optimization}--B1. Here, $C_1, C_2, \ldots, C_{24}$ denote the single-qubit Clifford gates and refer specifically to the representation of these gates in \cite{xia2015randomized} and \cite{ball2016effect}.}
\begin{center}
\begin{tabular}{ccccc} \toprule
    {Gate} & {$\probability{\Lambda=I\mid C_i}$} & {$\probability{\Lambda=X\mid C_i}$} & {$\probability{\Lambda=Y\mid C_i}$} & {$\probability{\Lambda=Z\mid C_i}$} \\ \midrule
    $ C_1 $  & $ 0.990 $ & $0.00\dot{3}$ & $0.00\dot{3}$ & $0.00\dot{3}$ \\
       $ C_2 $  & $ 0.965 $ & $0.012\dot{3}$ & $0.010\dot{3}$ & $0.012\dot{3}$ \\
            $ C_3 $  & $ 0.983 $ & $0.004\dot{3}$ & $0.008\dot{3}$ & $0.004\dot{3}$ \\
               $ C_4 $  & $ 0.977 $ & $0.008\dot{3}$ & $0.010\dot{3}$ & $0.004\dot{3}$ \\
                    $ C_5 $  & $ 0.969 $ & $0.011\dot{3}$ & $0.007\dot{3}$ & $0.012\dot{3}$ \\
                       $ C_6 $  & $ 0.984 $ & $0.006\dot{3}$ & $0.004\dot{3}$ & $0.005\dot{3}$ \\
                            $ C_7 $  & $ 0.979 $ & $0.004\dot{3}$ & $0.01\dot{3}$ & $0.00\dot{3}$\\
                               $ C_8 $  & $ 0.987 $ & $0.004\dot{3}$ & $0.003\dot{3}$ & $0.005\dot{3}$ \\
                                $ C_9 $  & $ 0.979 $ & $0.00\dot{3}$ & $0.009\dot{3}$ & $0.008\dot{3}$ \\
                                    $ C_{10} $  & $ 0.985 $ & $0.005\dot{3}$ & $0.005\dot{3}$ & $0.004\dot{3}$ \\
                                      $ C_{11} $ & $ 0.980 $ & $0.007\dot{3}$ & $0.00\dot{3}$ & $0.009\dot{3}$ \\
        $ C_{12} $ & $ 0.975 $ & $0.008\dot{3}$ & $0.006\dot{3}$ & $0.010\dot{3}$ \\
           $ C_{13} $  & $ 0.974 $ & $0.011\dot{3}$ & $0.006\dot{3}$ & $0.008\dot{3}$ \\
                $ C_{14} $  & $ 0.975 $ & $0.007\dot{3}$ & $0.006\dot{3}$ & $0.011\dot{3}$ \\
                    $ C_{15} $  & $ 0.972 $ & $0.01\dot{3}$ & $0.009\dot{3}$ & $0.005\dot{3}$ \\
                        $ C_{16} $  & $ 0.980 $ & $0.004\dot{3}$ & $0.009\dot{3}$ & $0.006\dot{3}$ \\
                            $ C_{17} $  & $ 0.979 $ & $0.006\dot{3}$ & $0.009\dot{3}$ & $0.005\dot{3}$ \\
                                $ C_{18} $  & $ 0.982 $ & $0.010\dot{3}$ & $0.004\dot{3}$ & $0.00\dot{3}$ \\
                                    $ C_{19} $  & $ 0.977 $ & $0.006\dot{3}$ & $0.004\dot{3}$ & $0.012\dot{3}$ \\
                                        $ C_{20} $  & $ 0.976 $ & $0.011\dot{3}$ & $0.007\dot{3}$ & $0.010\dot{3}$ \\
$ C_{21} $  & $ 0.975 $ & $0.007\dot{3}$ & $0.07\dot{3}$ & $0.010\dot{3}$ \\
$ C_{22} $  & $ 0.967 $ & $0.007\dot{3}$ & $0.007\dot{3}$ & $0.010\dot{3}$\\
$ C_{23} $  & $ 0.974 $ & $0.01\dot{3}$ & $0.006\dot{3}$ & $0.006\dot{3}$ \\
$ C_{24} $  & $ 0.978 $ & $0.012\dot{3}$ & $0.005\dot{3}$ & $0.004\dot{3}$ \\
 \bottomrule
\end{tabular}
\label{table:Specific_error_probabilities}
\end{center}
\end{table}

\section{Method to find all reachable stabilizer states}\label{appendix1}

All reachable stabilizer states can be found given the finite unitary group $ \mathcal{G}_n $ of gates (and noise) and the initial stabilizer state $ \ket{\psi_0} $. Given an initial stabilizer state $ \ket{\psi_0} $, the reduced states can be found by the following steps. First list all gates (and noise) $ \lbrace \mathcal{M}_1, \mathcal{M}_2, \cdots, \mathcal{M}_n\rbrace $ in group $ \mathcal{G}_n $. All reachable states are then $ \lbrace \mathcal{M}_1\ket{\psi_0}, \mathcal{M}_2\ket{\psi_0}, \cdots, \mathcal{M}_n\ket{\psi_0}\rbrace $. At last, any two states $  \mathcal{M}_i\ket{\psi_0} $ and $  \mathcal{M}_j\ket{\psi_0} $ that satisfies $ \mathcal{M}_i\ket{\psi_{0}}=e^{i\gamma} \mathcal{M}_j\ket{\psi_{0}}$ will fall into the same state.

\section{Pseudo-code for gate-limited simulated annealing}
\label{appendix:Pseudo_code_for_gate_limited_simulated_annealing}

In Algorithm~\ref{pseudo:Gate_limited_simulated_annealing}, we present the pseudo-code for the simulated annealing algorithm when restricting to a subset of available gates.

\begin{algorithm}[!hbtp]
\KwIn{A group $\mathcal{G}_n$, a set $ \mathcal{A}\subseteq\mathcal{G}_n $, a circuit $\{ U_1, \ldots, U_\tau \}$, and number of iterations $w$}
\KwOut{A revised circuit $\{ \itr{G}{w}_1, \ldots, \itr{G}{w}_\tau \}$}
\Begin{
Initialize $\{ \itr{G}{0}_1, \ldots, \itr{G}{0}_\tau \} = \{ U_1, \ldots, U_\tau \}$\;
\For{$\eta \leftarrow 1$ \KwTo $w$}{
Collect all $ m $ neighboring gates $ \{(\itr{G}{\eta}_1, \itr{G}{\eta}_{2}), \ldots, (\itr{G}{\eta}_{m-1}, \itr{G}{\eta}_{m})\} $ with at least one replaceable candidate neighboring gates $ \{ \itr{G}{\eta+1}_w\in\mathcal{A}, \itr{G}{\eta+1}_{w+1}\in\mathcal{A}\} $\;
Choose $I \in [m-1]$ uniformly at random\;
Replace $(\itr{G}{\eta}_I, \itr{G}{\eta}_{I+1})$ by any gate pair in $ \{(\tilde{G_1}, \tilde{G_2})\in \mathcal{A}^2 \mid \itr{G}{\eta}_{I+1} \itr{G}{\eta}_I = \tilde{G_2} \tilde{G_1} \}$ uniformly at random and then obtain the new circuit $ J $\;
Choose $X \in [0,1]$ uniformly at random\;
\uIf{$X \leq \alpha_{ \itr{G}{\eta}, J}(T_\eta)$}{
Set $\itr{G}{\eta+1} = J$\;
}
\Else{
Set $\itr{G}{\eta+1} = \itr{G}{\eta}$\;
}
}
}
\caption{Pseudo-code for gate-limited simulated annealing.}
\label{pseudo:Gate_limited_simulated_annealing}
\end{algorithm}

\section{Improved circuits}\label{optimized_circuit}

\begin{figure}[!hbtp]
\centering
\subfloat{
\includegraphics[
width=0.45\linewidth
]%
{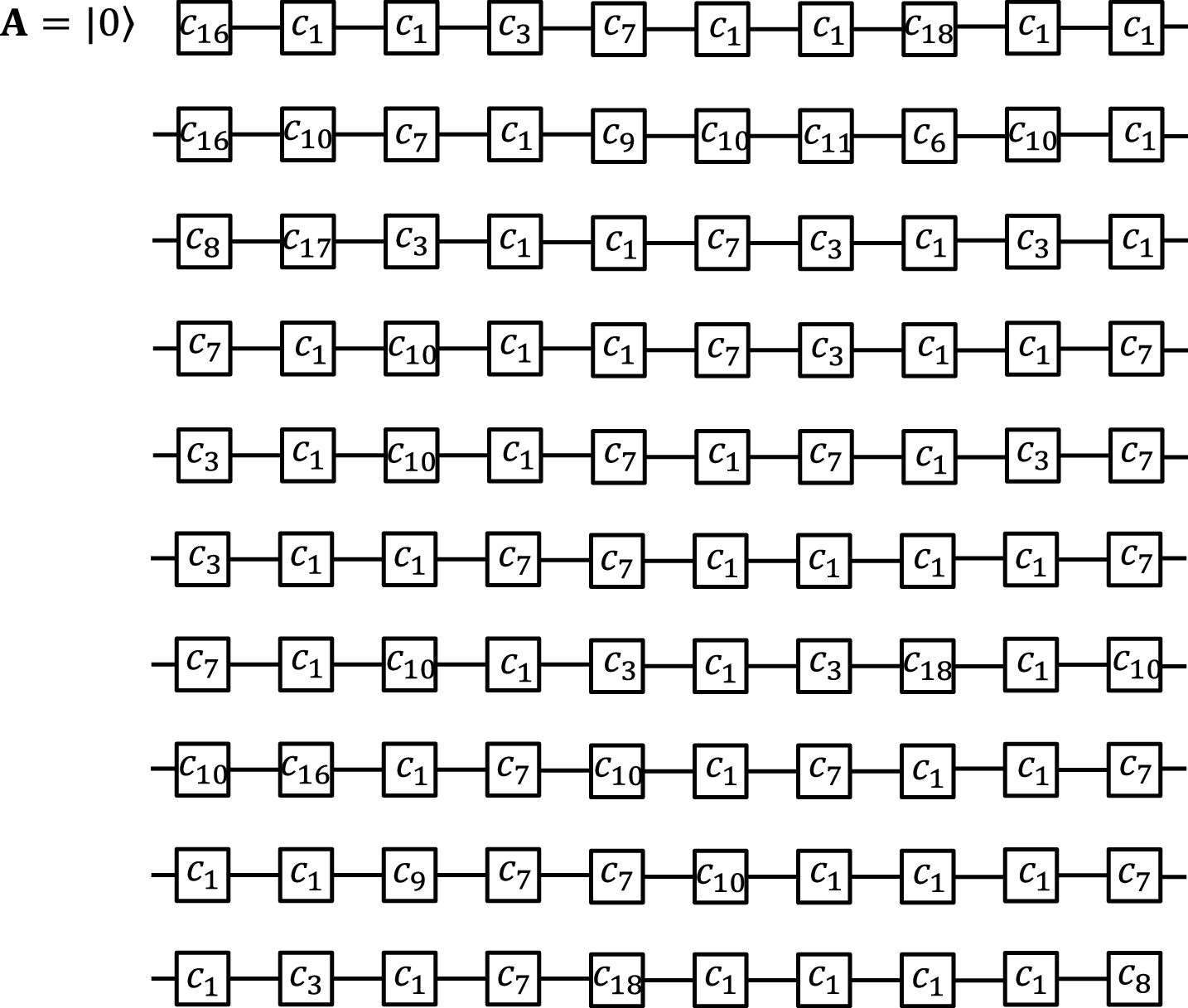}%
}
\,
\subfloat{
\includegraphics[
width=0.45\linewidth
]%
{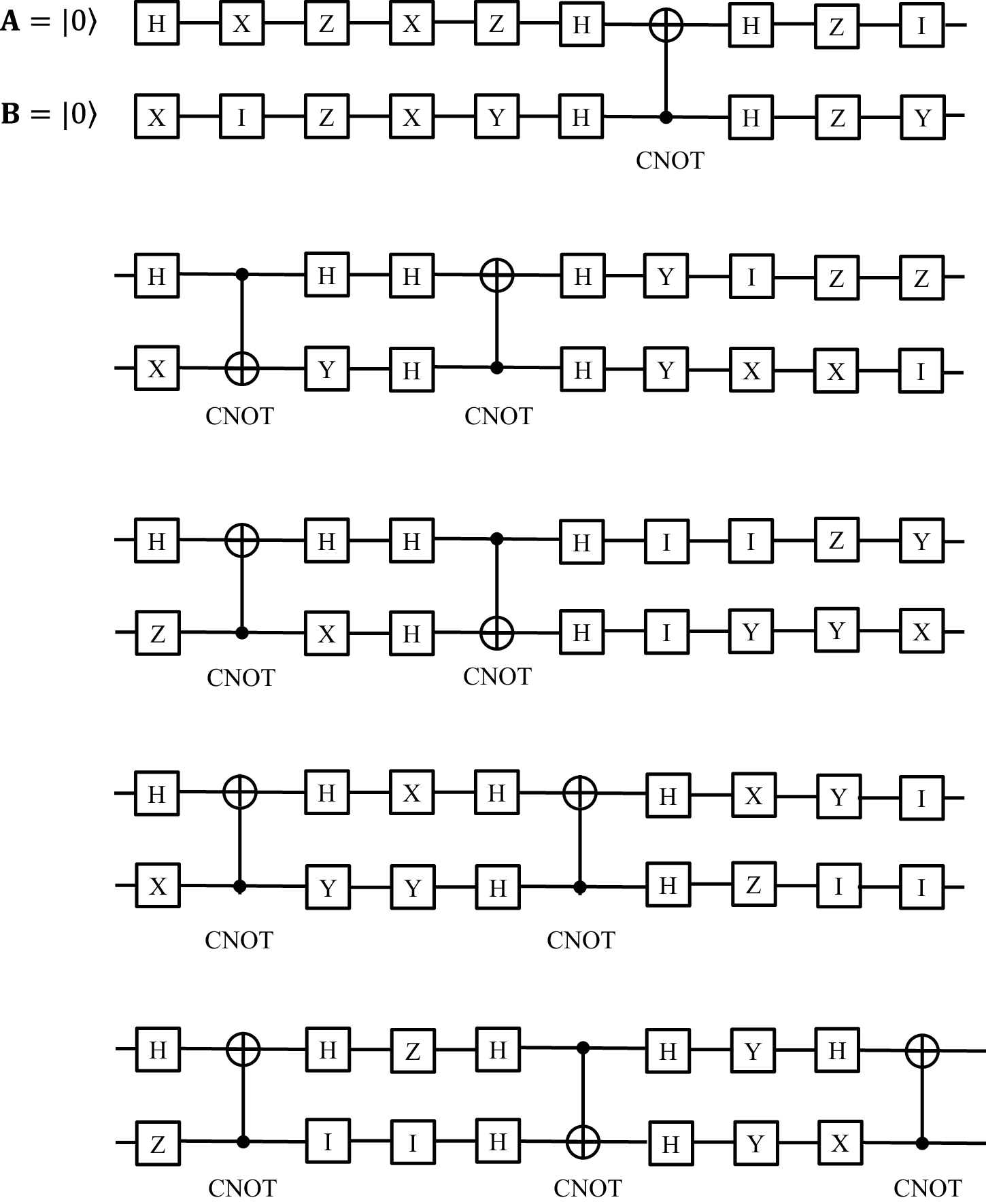}%
}
\caption{(left) The entire improved one-qubit circuit with circuit length $ \tau=100 $ obtained by Algorithm \ref{pseudo:The_Simulated_Annealing_algorithm} ($C=0.004$). (right) The entire improved two-qubit circuit with circuit length $ \tau=50 $ obtained by Algorithm \ref{pseudo:Gate_limited_simulated_annealing} ($C=0.002$).}
\label{fig:The_improved_circuits}
\end{figure}

In \refFigure{fig:The_improved_circuits}, we present the circuits with the lowest error accumulation rates found by our implementations of the two simulated annealing algorithms.

\section{Error distribution}
\label{appendix:Error_distribution}
Recall that $Q_{y,v}(t) = \probability{ Y_{t+1} = v | Y_t = y }$. In the experiment of \refFigure{fig:Heatmap_for_DeutschJozsas_Algorithm}, we assume a gate-dependent error model in which only single qubit errors occur that are (conditionally) i.i.d.\ on both qubits: that is
\begin{equation}
\probability{ \Lambda_{t+1} = \lambda | Y_t = y } 
= 
\begin{cases}
\zeta_{U_{t+1}}(\lambda_1) \zeta_{U_{t+1}}(\lambda_2) & \textrm{if } \lambda = ( \lambda_1, \lambda_2 ) \in \mathcal{C}_1^{\otimes 2} \\
0 & \textrm{otherwise,} \\
\end{cases}
\end{equation}
for a set of distributions $\{ \zeta_g | g \in \mathcal{C}_2 \}$, say. It now follows from the law of total probability that
\begin{equation}
Q_{y,v}(t) 
= \sum_{ ( \lambda_1, \lambda_2 ) \in \mathcal{C}_1^{\otimes 2} } \indicator{ ( \lambda_1 \otimes \lambda_2 ) U_{t+1} y \rho y^\dagger U_{t+1}^\dagger ( \lambda_1 \otimes \lambda_2 )^\dagger = v \rho_0 v^\dagger } \zeta_{U_{t+1}}(\lambda_1) \zeta_{U_{t+1}}(\lambda_2).
\end{equation}
Now, specifically, the error probabilities for e.g.\ the first qubit are set in the numerical experiment as shown in \refTable{table:Single_qubit_error_probabilities_for_the_DeutschJozsa_Algorithm_experiment}. The error probabilities for the second qubit are set similarly so. Note that \emph{not} applying a gate gives the lowest error rate; applying a single-qubit gate results in a medium error rate; and applying a two-qubit gate gives the largest probability that an error may occur.

\begin{table}[!hbtp]
\caption{\revised{The error probabilities for the first qubit as set in the numerical experiment that generates \refFigure{fig:Heatmap_for_DeutschJozsas_Algorithm}. The error probabilities for the second qubit are set similarly so.}}
\begin{tabular}{ccc}
\toprule
Case $g = I \times \mathcal{C}_1$: 
&
Case $g \in ( \mathcal{C}_1 \backslash I ) \times \mathcal{C}_1$:
&
Case $g \in \mathcal{C}_2 \backslash \mathcal{C}_1^{\otimes 2}$:
\\
\midrule
$
\zeta_g(\lambda_1) 
= 
\begin{cases}
0.990 & \textrm{if } \lambda_1 = I, \\
0.006 & \textrm{if } \lambda_1 = X, \\
0.003 & \textrm{if } \lambda_1 = Y, \\
0.001 & \textrm{if } \lambda_1 = Z, \\
0 & \textrm{otherwise}. \\
\end{cases}
$
&
$
\zeta_g(\lambda_1) 
= 
\begin{cases}
0.950 & \textrm{if } \lambda_1 = I, \\
0.030 & \textrm{if } \lambda_1 = X, \\
0.015 & \textrm{if } \lambda_1 = Y, \\
0.005 & \textrm{if } \lambda_1 = Z, \\
0 & \textrm{otherwise}. \\
\end{cases}
$
&
$
\zeta_g(\lambda_1) 
= 
\begin{cases}
0.900 & \textrm{if } \lambda_1 = I, \\
0.060 & \textrm{if } \lambda_1 = X, \\
0.030 & \textrm{if } \lambda_1 = Y, \\
0.010 & \textrm{if } \lambda_1 = Z, \\
0 & \textrm{otherwise}. \\
\end{cases}
$
\\
\bottomrule
\end{tabular}
\label{table:Single_qubit_error_probabilities_for_the_DeutschJozsa_Algorithm_experiment}
\end{table}
\vspace{-1em}

\section{Pseudo-code for the simulated annealing algorithm that improves the quantum circuit that implements Deutsch--Jozsa's Algorithm}
\label{appendix:simulated_annealing_3}
The algorithm that was used to generate the improved circuits for \refFigure{fig:Heatmap_for_DeutschJozsas_Algorithm} is shown in Algorithm~\ref{pseudo:The_Simulated_Annealing_algorithm_for_improving_DeutschJozsasAlgorithm}.

\begin{algorithm}[!hbtp]
\footnotesize
\KwIn{A circuit $\{ U_1, \ldots, U_\tau \}$ with $U_1, \ldots, U_\tau \in \mathcal{C}_2$, and number of iterations $w$}
\KwOut{A revised circuit $\{ \itr{G}{w}_1, \ldots, \itr{G}{w}_\tau \}$}
\Begin{
Initialize $\{ \itr{G}{0}_1, \ldots, \itr{G}{0}_\tau \} = \{ U_1, \ldots, U_\tau \}$\;
\For{$\eta \leftarrow 1$ \KwTo $w$}{
Choose $I \in [\tau-1]$ uniformly at random\;
Choose $B \in \{ -1, +1 \}$ uniformly at random\;
Choose $G \in \mathcal{C}_1 \otimes \mathcal{C}_1$ uniformly at random\;
\uIf{$B = -1$}{
Set $J_{I} = G, J_{I+1} = \itr{G}{\eta}_{I+1} \itr{G}{\eta}_{I} G^\gets, J_i = \itr{G}{\eta}_i \, \forall_{i \neq I, I+1}$\;
}
\Else{
Set $J_{I+1} = G, J_{I} = G^\gets \itr{G}{\eta}_{I+1} \itr{G}{\eta}_{I}, J_i = \itr{G}{\eta}_i \, \forall_{i \neq I, I+1}$\;
}
Choose $X \in [0,1]$ uniformly at random\;
\uIf{$X \leq \exp{ \bigl( - ( {1} / {T_\eta} ) \max{ \{ 0, u(J) - u(\itr{G}{\eta}) \} } \bigr) }$}{
Set $\itr{G}{\eta+1} = J$\;
}
\Else{
Set $\itr{G}{\eta+1} = \itr{G}{\eta}$\;
}
}
}
\caption{Pseudo-code for the simulated annealing algorithm that improves the quantum circuit that implements Deutsch--Jozsa's Algorithm for one classical bit.}
\label{pseudo:The_Simulated_Annealing_algorithm_for_improving_DeutschJozsasAlgorithm}
\end{algorithm}
\end{document}